\documentclass{WileyMSP-template}
\usepackage{graphicx}% Include figure files
\usepackage{dcolumn}% Align table columns on decimal point

\usepackage{bm}% bold math
\usepackage{xcolor}
\usepackage{mathtools}
\usepackage{verbatim}
\usepackage{amssymb}

\usepackage{physics}
\usepackage[normalem]{ulem}
\usepackage{lineno}
\usepackage{cite}

\usepackage{hyperref}
\hypersetup{colorlinks=true, citecolor=blue}

\begin{document}

%\pagestyle{fancy}
%\rhead{\includegraphics[width=2.5cm]{vch-logo.png}}

\title{Anomalous Loss Reduction Below Two-Level System Saturation in Aluminum Superconducting Resonators}

\maketitle

% Author: Please give full first and last names for authors and include * after the name of all corresponding authors

\author{Tamin Tai}
\author{Jingnan Cai}
\author{Steven M. Anlage**}

% Affiliations: Please provide academic titles (Prof. or Dr.) for all authors where applicable, and include an institutional email address for all corresponding authors
\begin{affiliations}
Dr. Tamin Tai, Jingnan Cai, Prof. Steven M. Anlage\\
Quantum Materials Center, Department of Physics, University of Maryland, College Park, MD 20742-4111, USA\\
Email Address: anlage@umd.edu

\end{affiliations}

% Keywords: Please provide a minimum of three and a maximum of seven keywords, separated by commas


\begin{abstract}
Superconducting resonators are widely used in many applications such as qubit readout for quantum computing, and kinetic inductance detectors. These resonators are susceptible to numerous loss and noise mechanisms, especially the dissipation due to two-level systems (TLS) which become the dominant source of loss in the few-photon and low temperature regime. In this study, capacitively-coupled aluminum half-wavelength coplanar waveguide resonators are investigated. Surprisingly, the loss of the resonators was observed to decrease with a lowering temperature at low excitation powers and temperatures below the TLS saturation. This behavior is attributed to the reduction of the TLS resonant response bandwidth with decreasing temperature and power to below the detuning between the TLS and the resonant photon frequency in a discrete ensemble of TLS. When response bandwidths of TLS are smaller than their detunings from the resonance, the resonant response and thus the loss is reduced. At higher excitation powers, the loss follows a logarithmic power dependence, consistent with predictions from the generalized tunneling model (GTM). A model combining the discrete TLS ensemble with the GTM is proposed and matches the temperature and power dependence of the measured internal loss of the resonator with reasonable  parameters.

\end{abstract}

\section{\label{sec:level1}Introduction}

Two-dimensional (2D) planar high internal quality factor ($Q_\text{i}$) superconducting resonators have been widely fabricated  and investigated in recent times for applications such as single photon detectors,\cite{Zmuidz12} kinetic inductance detectors,\cite{Baselmans} and quantum buses in quantum computing technology\cite{Krantz}. Tremendous progress has been made in terms of design, fabrication and measurement techniques, which has led to orders of magnitude increase in coherence time and improved quantum fidelity of the quantum gates \cite{Krantz,Kjaergaard,Hutchings}. In microwave measurements, although all qubits are operated at an excitation frequency well below the superconducting gap energy, microwave photons can be absorbed by quasiparticles, which in turn interact with the phonon bath, creating non-equilibrium distributions of both quasiparticles and phonons\cite{ChangScal77,Goldie,Visser2014,Budoyo2015}. This process affects the population of quasiparticles, in addition to pair-breaking process induced by cosmic rays,\cite{Vep20} higher order microwave harmonics, and stray infrared radiation \cite{Visser2011,Barends11,Rangga}. These non-equilibrium quasiparticles are one limiting factor on superconducting resonator $Q_\text{i}$ and qubit coherence, which can reduce both the qubit relaxation time $(T_1^{Qubit})$ and the coherence time $(T_2^{Qubit})$\cite{de_Leon}.\\ 

Another comparable loss mechanism due to two-level systems (TLS) is also ubiquitous in 2D superconducting resonators \cite{And72,Phil72,Schick77,Bla77,Black78_supporting,Phillips87,Martinis05,Gao07,Gao08,Ocon08,Kumar08,Barends08,Macha10,Burin18,Muller19,Yu21}. Despite the elusive microscopic origin of the TLS (some recent works suggesting hydrogen impurities in alumina as one candidate for TLS \cite{Hold13_supporting,Gordon}), TLS can be simply modeled as electric dipoles that couple to the microwave electric field. In general, TLS are abundant in amorphous solids and can also exist in the local defects of crystalline materials. They are found in three kinds of interfaces in the superconducting resonators: the metal-vacuum interface due to surface oxide or contaminants; the metal-dielectric substrate interface due to residual resist chemicals and buried adsorbates; and the dielectric substrate-vacuum interface with hydroxide dangling bonds, processing residuals, and adsorbates\cite{Oliver13}. To address these issues, different kinds of geometry of coplanar waveguide (CPW) structure have been proposed and fabricated, with more care given to the surface treatment to alleviate the TLS losses \cite{Bruno}. For example, a trenched structure in the CPW helps to mitigate the metal-dielectric TLS interaction with the resonator fields\cite{Calusine,Melville}. These efforts have improved the 2D resonator intrinsic quality factor to more than 1 million in recent realizations of high-$Q_\text{i}$ resonators \cite{Calusine,Jeremy_Sage,Chiaro,Richardson,Melville}.  Nevertheless, TLS still exist even in extremely high $Q_\text{i}$ 3D superconducting radio frequency cavities used in particle accelerator applications\cite{Romanenko}. Recently, other sources of TLS loss have been proposed based on quasiparticles trapped near the surface of a superconductor\cite{Graaf20}.

Clearly TLS loss is a universal issue in superconducting resonators. However, at microwave frequencies, this loss was long thought to be constant under low microwave power and low temperature below TLS saturation \cite{Phil72,Jack72,Hunk76,Bla77,Schick77,Burin18}.  Measurements in this regime were limited due to the constraints of noise levels in both electronic equipment and the thermal environment. Therefore, experimental investigation of TLS at low temperatures and microwave excitation are important, and would assist the superconducting quantum information community to understand its effect on operating quantum devices. 

\begin{figure*}
\includegraphics[width=180mm]{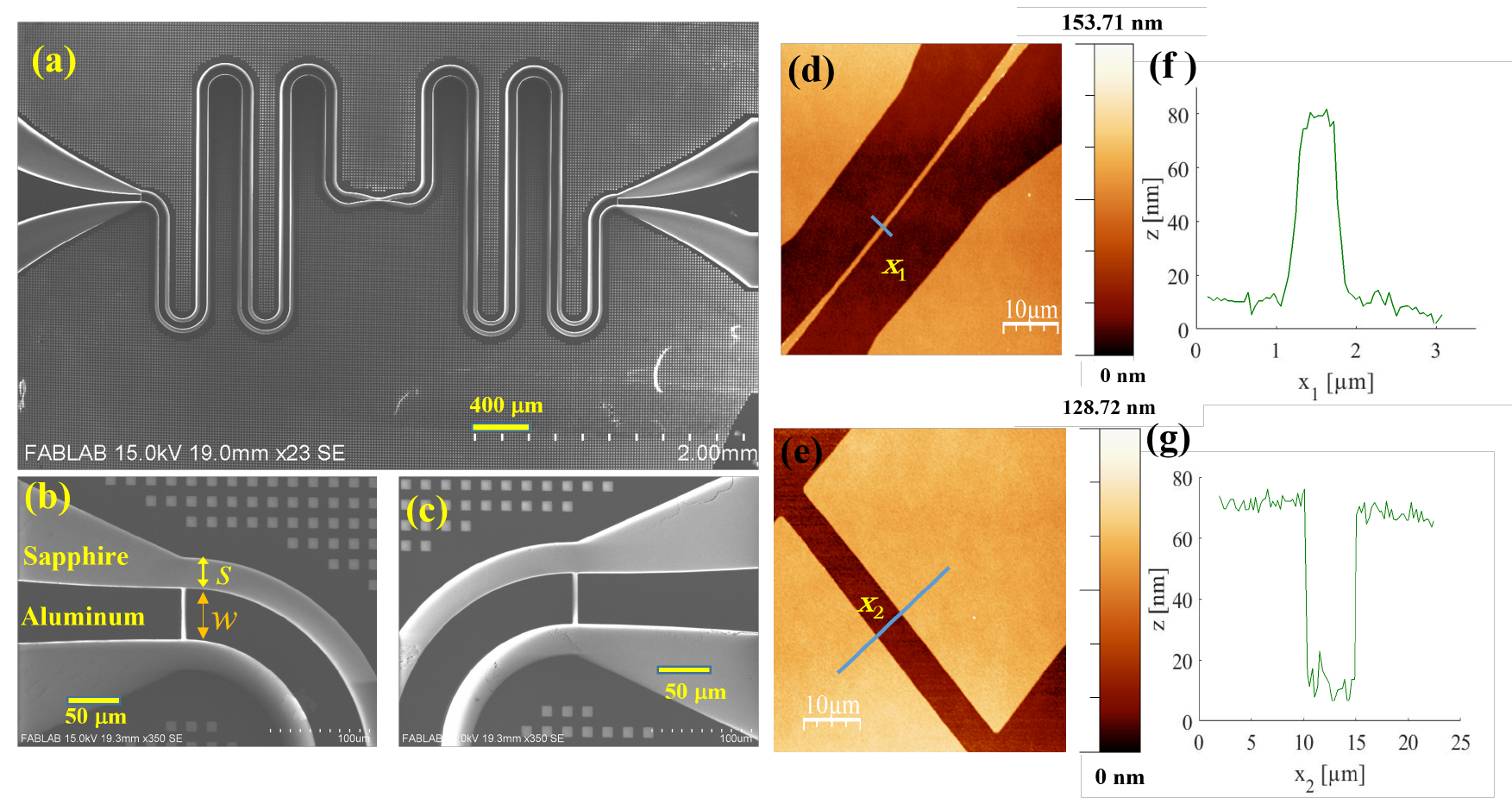} 
\centering
\caption{\label{fig:Device}{An SEM image of the aluminum CPW resonator on a sapphire substrate.  (b) (c) Zoom-in SEM images of the left and right capacitive couplers. (d) AFM image highlighting the tapered center conductor with a 1 $\mu m$ wide center trace near the center of the resonator, and (e) AFM topography image highlighting the 5 $ \mu m$ wide capacitive coupler from (b) or (c). Note that the AFM probe scanning direction is 45 degrees with respect to the center-line direction to reduce AFM scanning artefacts.} (f) Line scan profile of AFM image to show thickness of the center line in (d). (g) Line scan profile of the capacitive coupler. Both line scans show an Al film thickness of 70 nm.}
\end{figure*}

We have designed a 2D half wavelength resonator with a tapering geometry that gradually shrinks the signal line width $w$ from 50 $\mu m$ down to 1 $\mu m$ at the center where many three-junction flux qubits could be hosted and strongly coupled for the study of the collective behavior of quantum meta-materials. Analogous to cavity quantum electrodynamics, qubits serve as artificial meta-atoms with mutual coupling \cite{Du,Rak08,Jung14,Zag16,Laz18} and can be read out through the dispersive frequency shift of the cavity\cite{You,Macha14,Shulga18}.  Theoretical publications discussing the physics of qubit arrays coupled to the harmonic cavities predict a number of novel collective behaviors of these meta-atoms \cite{Xiang,Mukhin13,Volkov14}. In this paper, we report our finding on the TLS loss in the low power and low temperature limit of this particular design of capacitively-coupled half-wavelength resonator, without the qubits. The technique of very low power microwave measurement with low noise to enhance the signal-to-noise ratio (SNR) is critical for measuring this TLS behavior. 

\section{Experimental Methods}
Aluminum (Al) half-wavelength $(\lambda/2)$ CPW resonators on sapphire substrates were designed with a center line width $w=50$ $\mu m$ and spacing $s=30$ $\mu m$ (the distance between center conductor line and ground plane as illustrated in Fig. \ref{fig:Device}(b) to maintain the  characteristic impedance near 50 $\Omega$ in the meander part. At the center of the resonator a tapering structure narrows the center line width down to $w=1 \: \mu m$ and spacing to $s=12 \: \mu m$, which gradually increases the characteristic impedance to 100 $\Omega$ at the resonator center.  Fig. \ref{fig:Device} (a) shows a perspective view of the resonator in a diced chip with a designed fundamental frequency around 3.6 GHz.  The entire resonator is surrounded by many 10 $ \mu m$ by 10 $\mu m$ vortex moats. The resonator is symmetric and capacitively coupled through 5 $\mu m$ gaps (Fig. \ref{fig:Device} (b) and \ref{fig:Device} (c)) in the center conductor. A topographic image of the narrowed resonator center section is shown in Fig. \ref{fig:Device} (d) with a critical dimension around $w=$ 1 $ \mu m$ in width.  Line cuts shown in the AFM images in Fig. \ref{fig:Device} (d), (e) show that the Al film is 70 nm thick.   

This CPW resonator was fabricated using standard photo-lithography procedures. First, a 70 nm thick Al film was deposited on a 3-inch diameter sapphire wafer using thermal evaporation technology with a background pressure of $\sim 3\times 10^{-7}$ mbar. Then a thin SHIPLEY1813 photo-resist was coated on top of the film and exposed to UV through the designed photomask. The UV exposed wafer was developed and then wet etched by commercial Transene Aluminum Etchant. The remaining photoresist was stripped off by acetone and the entire wafer was cleaned by methanol and isopropanol.  Finally, the wafer was coated in a protective photo-resist and then diced into many chips. After dicing, the protective photo-resist was removed and the chip was mounted on a printed circuit board bolted inside a copper box. Several lumps of indium were pressed between the on-chip ground planes and the copper box ground to achieve a continuous ground contact, which mitigates parasitic resonant microwave modes due to uneven electrical grounding. The indium lumps also secured the chip in the center of the printed circuit board. The on-chip transmission line is wire-bonded to the center conductor of the transmission line on the printed circuit board by gold wires. Finally, the copper box is capped by a copper lid to eliminate stray light illumination.  

The device was placed in a closed Cryoperm cylinder in a BlueFors (BF-XLD 400) croygen-free dilution refrigerator (base temperature 10 mK) to minimize any stray DC magnetic field, and the shield was thermally anchored to the mixing chamber plate.  The microwave excitation was attenuated by a series of attenuators in the input line at different cooling stages in the dilution fridge before going into the resonator to reduce the noise. The transmitted signal was amplified twice through a cryogenic amplifier and a room temperature amplifier before being measured by a Keysight N5242A vector network analyzer (VNA). The low power measurements were performed using the smallest intermediate frequency bandwidth (1 Hz) of the VNA, with a 400 kHz span across the resonance, following 5 averages to reduce the random noise. A thru calibration of the setup was performed in a separate cool down to determine the overall loss/gain in the transmission lines leading to the resonator. Further details of the experimental setup for the high SNR measurement at very low microwave power can be found in section \ref{support Info_I} of the Supplemental Material.

\section{Experimental Data}

The measured transmitted signal $(S_{21}(f))$ has a fundamental $(\lambda/2)$ resonance peak around $f=$ 3.644 GHz at the fridge base temperature when sweeping the frequency, $f$.  The complex $S_{21}(f)$ signal is fitted to an equivalent circuit model of a two-port resonator capacitively coupled to external microwave excitation \cite{Pet98,Budoyo2015}. 

\begin{eqnarray}
S_{21}(f)=|S_\text{21,in}||S_\text{21,out}| \left( \frac{Q_\text{L}/Q_\text{c}}{1+2iQ_\text{L}(\frac{f}{f_0}-1)} e^{i\phi}\right)+C_0 
\label{eq:S21}
\end{eqnarray}
where $|S_{21,\text{in}}|$ and $|S_{21,\text{out}}|$ are the net loss or gain in the transmission of the input and output line, respectively. $Q_\text{L}$ is the loaded quality factor. $Q_\text{c}$ is the coupling quality factor representing the dissipation to the external circuit, $i=\sqrt{-1}$, $f_0$ is the resonance frequency of the half-wavelength $(\lambda/2)$ CPW resonator, $\phi$ is the phase and $C_0$ is an offset in the complex $S_{21}$ plane due to background contributions.\cite{Pet98} The internal quality factor, $Q_\text{i}$, inversely proportional to the the internal loss, $\delta=Q_\text{i}^{-1}$, is extracted from the identity $1/Q_\text{L} \equiv 1/Q_\text{i}+1/Q_\text{c}$. The absorbed power $P_\text{ab}$ of the resonator is characterized by the average number of circulating microwave photons in the cavity on resonance, which can be estimated using the approximation \cite{Budoyo2015,Weber} $\langle n\rangle=\frac{2Q_\text{L}^2 P_\text{in}}{Q_\text{c} \hbar \omega_0^2 }$ for a two-port device, where $\hbar$ is the reduced Planck constant, and $\omega_0=2\pi f_0$ is the angular frequency of the resonance.

\begin{figure*}[hbt!]
\includegraphics[width=200mm]{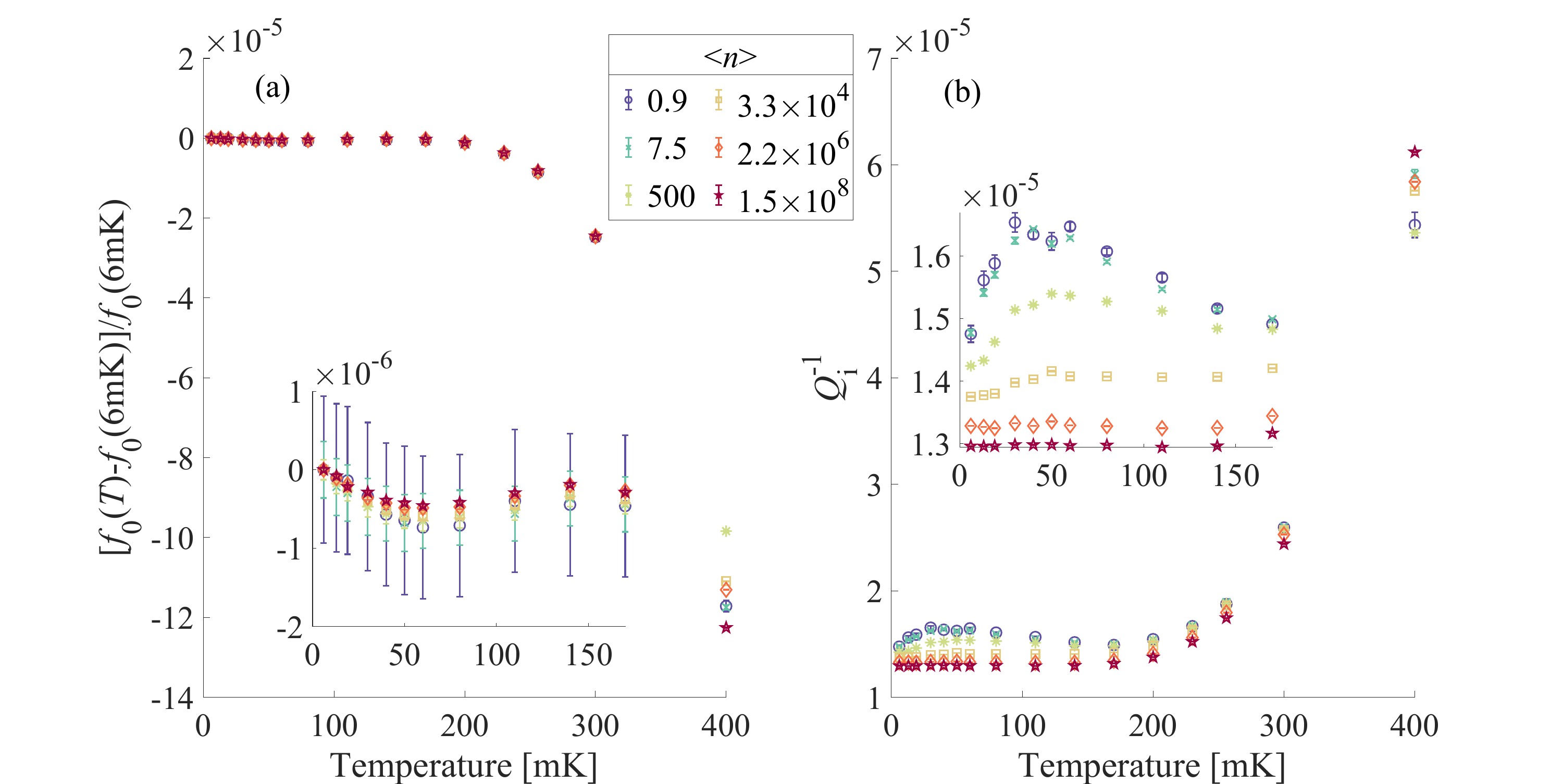}
\caption{\label{fig:f0_Qi} (a) Temperature dependent first harmonic resonant frequency shift $\Delta f/f_0(6 \text{mK})$, with $\Delta f = f_0 - f_0(6 \text{mK})$ of the $\lambda/2 $ Aluminum co-planar waveguide resonator on sapphire substrate measured at different excitation powers (average photon numbers).  Here $f_0(6 \text{mK})$ is the resonance frequency measured at the base temperature for each excitation power . (b) Temperature dependent loss (inverse of intrinsic quality factor, $Q_\text{i}^{-1}$) at its first harmonic frequency of a Aluminum co-planar waveguide resonator on sapphire substrate measured at different circulating photon numbers $\langle n \rangle$. Some of the error bars are smaller than the data point such as those for the high power and temperature measurements. }
\end{figure*}

Figure \ref{fig:f0_Qi}(a) illustrates the temperature dependence of the fractional resonant frequency shift from the resonance frequency at lowest temperature, $(f_0(T) - f_0(6 \text{mK}))/f_0(6 \text{mK})$, for different circulating microwave photon numbers inside the CPW resonator, where 6 mK is the measured fridge base temperature. The resonance frequencies start at their maxima at the fridge base temperature and then show local minima around 60 mK. This phenomenon seems to be independent of the average circulating photon number and can be explained by the standard tunneling model of TLS \cite{Hunk76}. Upon further increasing the temperature above 150 mK, the resonance frequencies quickly decrease due to the thermal quasiparticles, which increases the real and imaginary parts of the surface impedance of the superconducting resonator.  The inset focuses on the low temperature regime and shows a very small power dependence that is qualitatively similar to the strong field correction to the frequency shift in STM proposed by Gao, which  predicts smaller frequency shifts for higher power \cite{Gao_phdthesis}. 

The temperature dependence of the measured internal loss is shown in Fig. \ref{fig:f0_Qi}(b). For high power measurements ($\langle n \rangle > 10^6$),  the loss is constant at low temperatures (below 150 mK) which is expected for the typical non-interacting TLS. At higher temperatures, the loss increases due to thermal quasiparticles. For low power measurements ($\langle n \rangle < 10^6$), starting from the minimum temperature, the loss has an unusual increase at low temperatures, from the base temperature to a peak at 40 mK. The loss then drops with increasing temperature following the equilibrium value of the population difference in TLS \cite{Phillips87,Mcrae}.   Similar to the high power measurements, the loss rises again above 150 mK due to thermal quasiparticles. The observed loss decrease with decreasing temperature from 40 mK to 10 mK has not been explicitly acknowledged and discussed in prior work of microwave superconducting resonators until recently \cite{crowley_disentangling_2023}. Indications of an upturn in $Q_\text{i}(T)$ has been attributed to poor SNR and therefore treated as not statistically significant\cite{Mcrae,Burnett}.

\section{Modeling}
\subsection{Frequency shifts}
The power and temperature dependent frequency shifts have primarily been attributed to TLS and the dynamics of quasiparticles. These two  mechanisms could overlap and become difficult to distinguish in the operation of many superconducting devices, including resonators and qubits, \cite{de_Leon}. A simple model that combines both quasiparticles and TLS contribution in one equation describes the resonance frequency $\Delta f$ data in Fig. \ref{fig:f0_Qi},\cite{Kumar08,Bruno,gao_equivalence_2008}  

\begin{align}
 \frac{f_0(T)-f_0(0)}{f_0(0)}
 =&\frac{\delta_0}{\pi} 
 \left ( 
 \text{Re} \left[ \Psi (\frac{1}{2}+\frac{\hbar \omega}{2 \pi i k_\text{B} T}) \right ] -log (\frac{\hbar \omega }{2 \pi k_\text{B} T} ) 
 \right ) \nonumber \\
&- \frac{\alpha}{2} \left( \frac{n_\text{qp}}{2N_0\Delta_{\text{S}0}} \left[ 1+\sqrt{\frac{2\Delta_{\text{S}0}}{\pi k_\text{B} T}}\exp (\zeta)I_0(\zeta ) \right] \right) \label{TLS:Freq} 
\end{align}
where $\zeta = \frac{hf_0}{2k_\text{B} T}$, $f_0$ is the resonance frequency as a function of the temperature, $\delta_0$ is the zero temperature and zero power loss tangent from the TLS, $\Psi(\cdot)$ is the digamma function, $\alpha=L_\text{kinetic}/L_\text{total}$ is the kinetic inductance fraction of the CPW resonator, $N_0$ is the single spin density of states, $\Delta_{\text{S}0}$ is the aluminum superconducting gap at zero temperature, and $I_0(\cdot)$ is the 0th order modified Bessel function of the first kind. The first term in Eq. \ref{TLS:Freq} represents the frequency shift caused by the TLS mechanism \cite{Phillips87,Gao08} and the second term is the frequency shift due to quasiparticles using the Bardeen-Cooper-Schrieffer (BCS) model for $k_\text{B} T,\  hf_0  \ll \Delta_{\text{S}0}$, and written explicitly in terms of quasiparticle number density $n_\text{qp}$ \cite{gao_equivalence_2008}, including both thermal  and non-equilibrium quasiparticles. However, the model with only thermal quasiparticle $n_\text{th}=2 N_0 \sqrt{2 \pi k_\text{B} T \Delta_{\text{S}0}}\exp \left(-\frac{\Delta_{\text{S}0}}{k_\text{B} T } \right)$ (valid for $T\ll T_\text{c}$) seems to match the measurement sufficiently well, where $N_0 = 10^{47} $ $J^{-1}  \: m^{-3} \approx 1.74 \times 10^4\mu eV^{-1} \mu m^{-3}$ is the single spin density of states at the Fermi level \cite{Goldie,Rangga}.

\begin{figure*}[hbt!]
\includegraphics[width=100mm]{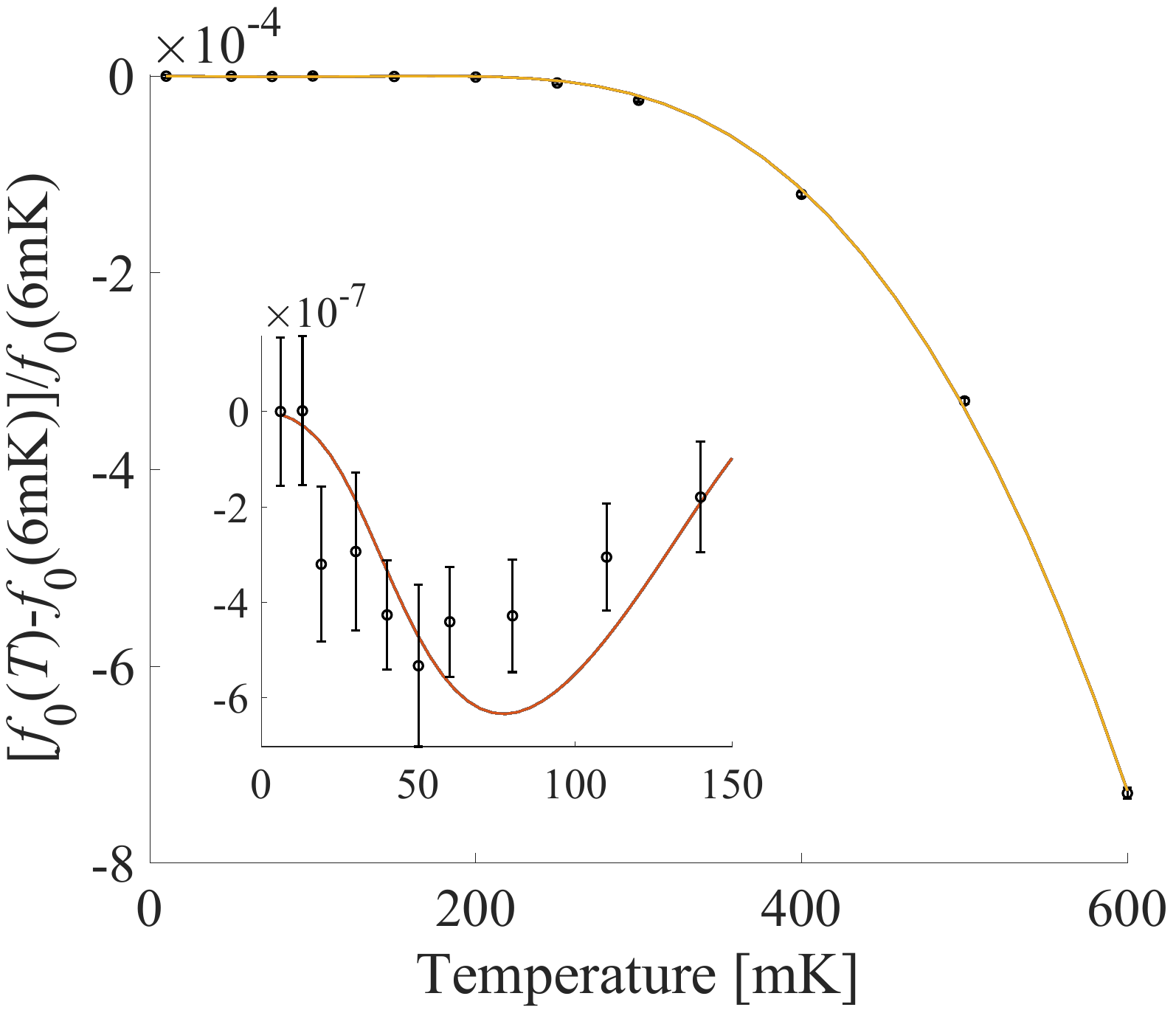}
\centering
\caption{\label{freq_T} Temperature dependent fundamental ($\lambda/2$) mode resonant frequency $f_0(T)$ of the Al CPW resonator on sapphire substrates at an external microwave excitation creating around one circulating photon. The inset highlights the low temperature regime where the frequency shift is dominated by the TLS mechanism. The dots are experimental data and solid line is the model fit to Eq. (\ref{TLS:Freq}).}  
\end{figure*}

The fit to the frequency shift data is shown in Fig. \ref{freq_T}, and the extracted fitting parameters indicate that the aluminum superconducting gap at zero temperature is $\Delta_{\text{S}0} \sim$170 $\mu eV$, a value close to the BCS gap approximation which is $1.76 k_\text{B} T_\text{c}$ with transition temperature $T_\text{c}=1.12 \: K$.  The values of the other fitting parameters are  $\alpha \approx 0.014$, and $\delta_0=9.6\times10^{-6}$.  The values of $\alpha$ and $\delta_0$ are consistent with other results on a variety of similar superconducting resonators\cite{Gao08,Macha10,Pappas11,Bruno}. 
%Note that this fit predicts the maximum $Q_\text{max}=(\delta_0)^{-1}=1.26 \times 10^5$ at low temperature and low power, 23 \% larger \textcolor{red}{This number has changed!} than what is observed in Fig. \ref{Qi_T}.  However, this is consistent with previous observations that attribute the difference between predicted and measured $Q$-factors to the difference in bandwidth of TLS states contributing to $Q$ and to frequency shift\cite{Pappas11,Bruno}.  

\subsection{Internal loss}
\begin{figure*}[hbt!]
\includegraphics[width=150mm]{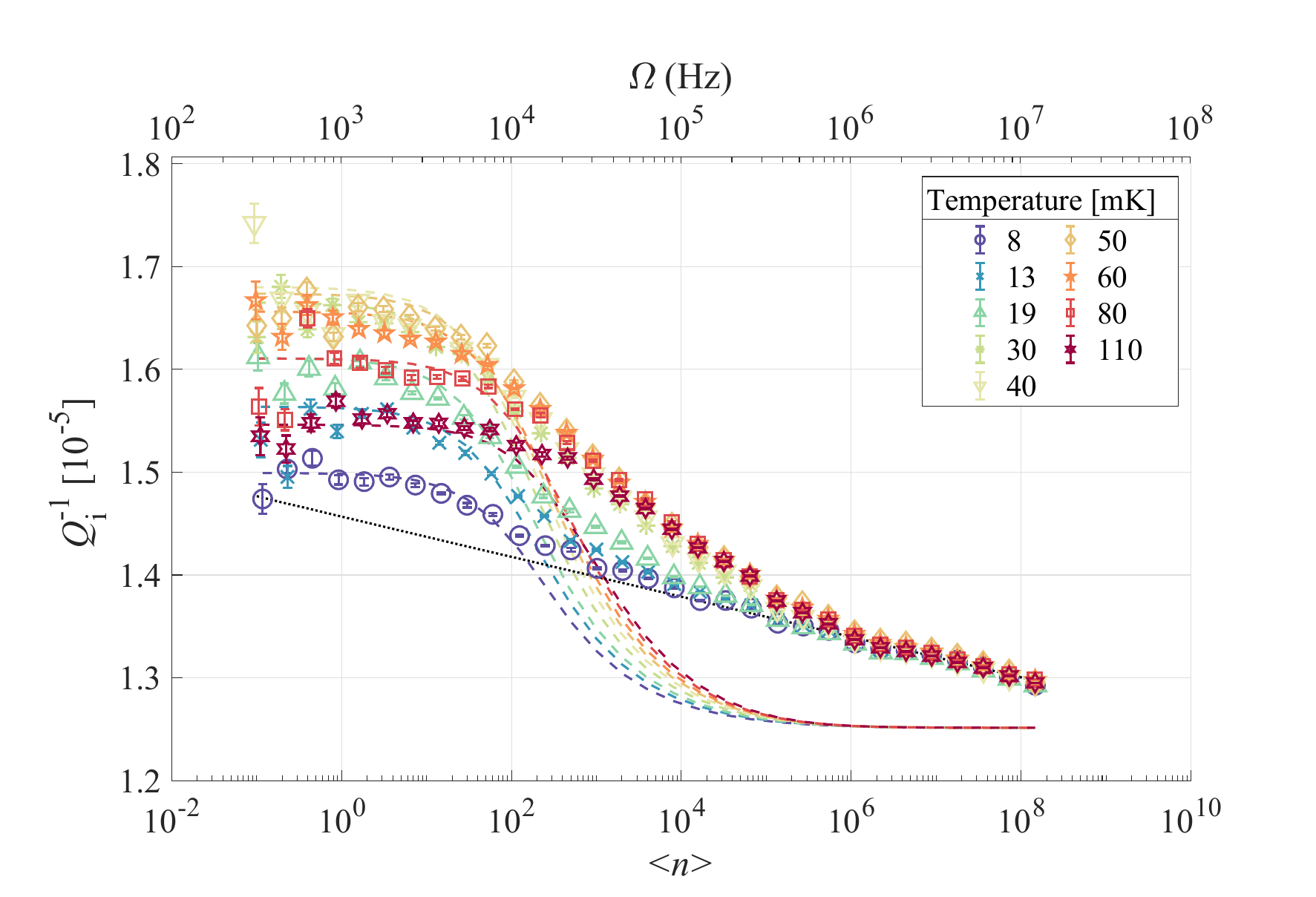}
\centering
\caption{\label{Qi_T} Internal loss $Q_\text{i}^{-1}$ as a function of power (measured by photon number $\langle n \rangle$ on lower axis, and Rabi frequency $\Omega$ on upper axis) at different temperatures for an Aluminum resonator on a sapphire substrate. The scatter plots are experimental data points, and the dashed lines are the fitting curves from the STM given in Eq. (\ref{STMloss}). There is a large deviation from STM power dependence at high power above TLS saturation power. The power dependence is more gradual than the STM prediction, and the loss has very weak temperature dependence, which resemble the logarithmic power dependence predicted by GTM. The black dotted line is the power dependence at high excitation power from GTM. \textcolor{black}{A constant background loss is assumed for all the fits.}}
\end{figure*}

Since the temperature dependent internal loss is dominated by the well-known thermal quasiparticles above 150 mK, this analysis focuses only on the low temperature data.  The power dependence of the loss $Q_\text{i}^{-1} (T)$ is shown in Fig. \ref{Qi_T} at different temperatures below the onset of thermal quasiparticle effects. Clearly, the loss shows a gradual power dependence above the low-power saturation, similar to previous experimental observations\cite{wang_improving_2009,Macha10,khalil_2011,Jeremy_Sage}, and is not consistent with  STM shown as the dashed curves.

%\sout{The slow power dependence is a clear signature of the generalized tunneling model (GTM) developed by Faoro and Ioffe, which predicts that TLS loss at high applied powers has a slow logarithmic dependence on power, shown as the black dotted line, and agrees with our data.} 

To account for the slower power dependence, many improvements on the STM have been proposed, such as introducing more than one species of TLS in the dielectrics, \cite{schechter_what_2008,khalil_evidence_2013,schechter_inversion_2013,schechter_nonuniversality_2018,yu_experimentally_2022} and accounting for the nonuniform field distribution in the resonator \cite{Ocon08}. In addition, there is another approach that generalizes the STM to \textcolor{black}{include a random telegraph noise on the TLS energy level due to strong interactions between a few TLS } \cite{Burn14,faoro_internal_2012,faoro_interacting_2015}, resulting in the generalized tunneling model (GTM) that can produce the logarithmic power dependence shown as the black dotted line in Fig. \ref{Qi_T}.\\
However, none of the existing models predicts a strong temperature dependence of loss below the TLS saturation. To interpret this unusual loss reduction in our aluminum resonators at low power and low temperature, we go beyond the assumption of a uniform distribution of TLS and invoke the discrete TLS contribution to the loss at low temperatures. A simple modification that sums over the discrete and detuned TLS responses near the resonance as in Eq. (\ref{discrete_sum})  is proposed. When combined with GTM, this model reproduces the full power and temperature dependence of the loss data: the gradual power dependence at high power as well as the observed anomalous temperature dependence of loss for $\langle n \rangle < 10^2$ and $T < 50$ mK. \textcolor{black}{It should be emphasized that the discrete TLS assumption is independent of GTM. Attempts to apply the discrete and detuned TLS formalism to the modified versions of STM are summarized in Supplementary Material Section \ref{other_model}.} To lay the foundations of the proposed model, \textcolor{black}{the following sections introduce key concepts of STM and GTM and derive several expressions used in the final model.}

\subsubsection{Conventional model for TLS loss}

The TLS formalism is based on a simple model for a single TLS that can be described by the Hamiltonian,
$H_\text{TLS}=\frac{1}{2}
\begin{pmatrix}
-\Delta   &  \Delta_0 \\
 \Delta_0 &  \Delta 
\end{pmatrix}$
where $\Delta $ is the asymmetry of the double well potential and $\Delta_0$ is the tunneling barrier energy between the  potential wells\cite{Phillips87}. A typical resonator hosts an ensemble of TLS with different values of $\Delta$ and $\Delta_0$ with their (assumed continuous) distribution function given as $ P(\Delta,\Delta_0)=P_0/ \Delta_0$, where $P_0\approx10^{44} J^{-1}m^{-3}$ is the density of states for TLS. \textcolor{black}{The distribution function is uniform in $\Delta$ in the conventional TLS model, but could take on a very weak dependence $\propto \Delta^\mu$ with $\mu\sim0.3$ for a system of very strongly interacting TLS, \cite{cuevas_density_1989,faoro_interacting_2015,churkin_anomalous_2021} such as the case assumed in GTM. For simplicity \textcolor{black}{and generality}, the following model uses the conventional distribution function, which is constant in $\Delta$. The fit with non-zero $\mu$ can be found in the Supplementary Material Section. \ref{alt_nonzero_mu}. }

The dynamics of a single TLS can be described by the linearized Bloch equations of the pseudospin $\vec S(t)$ (see Supplementary Material Section \ref{STM_sup} for details) that is characterized by the following four rates: Rabi frequency $\Omega\propto |\vec{E}|$, the frequency of the driving field $\omega$, the splitting between the two eigenenergies of TLS $\varepsilon=\sqrt{\Delta^2+\Delta_0^2}$, and the longitudinal and transverse relaxation rates of the TLS $\Gamma_{1,2}$ which are defined as 

\begin{align}
\Gamma_1 = \left(\frac{\Delta_0}{\varepsilon}\right)^2
\left[\frac{\gamma_\text{L}^2}{v_\text{L}^5 }+\frac{2\gamma_\text{T}^2}{v_\text{T}^5 } \right] 
\frac{\varepsilon^3}{2 \pi \rho \hbar^4 } &\coth (\frac{\varepsilon}{2 k_\text{B} T}) = \left(\frac{\Delta_0}{\varepsilon}\right)^2 \Gamma_1^\text{max} \quad\text{\cite{Jack72,Hunk76,Bla77,Gao_phdthesis}}\label{T1_eq}\\
&\Gamma_2=\Gamma_2^\text{ph}+\Gamma_\text{ds} \nonumber\\
\text{where}\quad\Gamma_2^\text{ph}=\Gamma_1/2 \quad\text{and}\quad &\Gamma_\text{ds}\sim 10^{-3}(k_\text{B} T/\varepsilon_\text{max})^\mu k_\text{B} T/\hbar \quad\text{\cite{jankowiak_origin_1993,faoro_interacting_2015}}
\label{Tds_eq}
\end{align}
Eq.(\ref{T1_eq}) describes the longitudinal relaxation rate dominated by the phonon process where $\gamma_\text{L}$ and $\gamma_\text{T}$ are the longitudinal and transverse deformation potentials, respectively, $v_\text{L}$ and $v_\text{T}$ are the longitudinal and transverse sound velocities, $\rho$ is the mass density, and $\Gamma_1^\text{max}$ is the maximum $\Gamma_1$ for the TLS with energy splitting $\varepsilon$, when $\Delta_0=\varepsilon$. Eq.(\ref{Tds_eq}) defines the transverse relaxation rate where $\Gamma_\text{ds}$ is the dephasing rate of the resonant TLS energy level $\varepsilon$, caused by its interactions with thermally activated TLS whose $\varepsilon \lesssim k_\text{B} T$, and is valid for low temperature measurement ($T<1$ K) \cite{jankowiak_origin_1993}. We note that \textcolor{black}{ $\mu=0$ for the conventional TLS model used here, and }$\Gamma_\text{ds} \sim 10^6\  \text{Hz}$ dominates over $\Gamma_1 \sim 10^3\ \text{Hz}$ in the typical cryogenic measurement of amorphous dielectrics \cite{faoro_interacting_2015}. Therefore, $\Gamma_2$ is often approximated as $\Gamma_\text{ds}$ and \textcolor{black}{is proportional to $T$.}  

In STM, the resonant dielectric response of a single TLS is expressed as \cite{And72,Schick77,faoro_internal_2012}:
\begin{equation}
\chi_\text{res}=\frac{m(\omega-\varepsilon/\hbar-i\Gamma_2)}{(\omega-\varepsilon/\hbar)^2+\Gamma_2^2(1+\Omega^2\Gamma_1^{-1}\Gamma_2^{-1})}
\label{singleTLS}
\end{equation}
where $m=\tanh (\varepsilon/(2k_\text{B} T))/2$ is the equilibrium value of $\langle S_z^0 \rangle$. The single TLS loss corresponds to the imaginary part of the response function in Eq.(\ref{singleTLS}) which is in the form of a Lorentzian in $\varepsilon/\hbar$ centered at $\omega$ with a width:
\begin{align}
    w=\Gamma_2\sqrt{1+\kappa}
    \label{lorentzian_width}
\end{align}
where $\kappa=\Omega^2\Gamma_1^{-1}\Gamma_2^{-1}$. For a typical TLS with $\varepsilon/h \approx$ 5 GHz and at reasonably low temperatures and powers, the width of its response $ w \approx\Gamma_2 \sim 1\ \text{MHz} \ll \omega$. Due to this sharp Lorentzian response function, the total loss is dominated by the resonant TLS whose energies $\varepsilon \sim \hbar\omega$.

The total dielectric loss is simply the integral of the single TLS contribution Eq.(\ref{singleTLS}) over the distribution of the TLS \cite{Schick77,Phillips87,Gao_phdthesis}.
\begin{equation}
\delta_\text{TLS}
=\frac{1}{\epsilon_\text{r}\epsilon_0}\int\int\int P(\varepsilon,\Delta_0)\left(\frac{\Delta_0 d_0}{\varepsilon}\right)^2\frac{\cos^2{\theta}}{\hbar}  
 \frac{m\Gamma_2}{\Gamma_2^2(1+\kappa)+(\varepsilon/\hbar-\omega)^2}\text{d}\varepsilon \text{d}\Delta_0 \text{d}\theta
\label{loss_int}
\end{equation}
where $\epsilon_\text{r}\epsilon_0$ is the permittivity of the host dielectric material, $P(\varepsilon,\Delta_0)$ is the distribution function of coherent TLS, obtained from $P(\Delta,\Delta_0)$ with a change of variable from $\Delta$ to $\varepsilon$, $d_0$ is the maximum transition electric dipole moment of the TLS with energy splitting $\varepsilon$, $\vec{E}$ is the applied microwave electric field on the TLS dipole, and $\theta$ is the angle between the applied electric field and the TLS dipole moment.

Evaluating this integral leads to the famous STM prediction of TLS loss \cite{Hunk76}
\begin{equation}
    \delta_\text{TLS}=\frac{\pi P_0d_0^2}{3\epsilon_\text{r}\epsilon_0}\frac{\tanh{\frac{\hbar\omega}{2k_\text{B} T}}}{\sqrt{1+(\Omega/\Omega_\text{c})^2}}
    \label{STMloss}
\end{equation}
where $\Omega_\text{c} \propto \sqrt{\Gamma_1^\text{max}\Gamma_2}$ is the critical Rabi frequency that characterizes the saturation of TLS. The loss is expected to have an inverse square root dependence on power after the TLS saturation, $\delta_\text{TLS}\sim \Omega^{-1} \propto P_\text{ab}^{-0.5}$ for $\Omega \gg \Omega_\text{c}$, which is much faster than observed in the data in Fig. \ref{Qi_T}. In fact, if one fits the data with a general power law \cite{Macha10} where the square root in the denominator of Eq.(\ref{STMloss}) is replaced by a fitting parameter, the resulting exponent is around -0.15, indeed a slower power dependence than predicted in STM.

\subsubsection{Effect of fluctuators on TLS loss}

The dephasing rate $\Gamma_\text{ds}$ introduced in Eq.(\ref{Tds_eq}) describes the \textcolor{black}{spectral diffusion} resulting from an average of weak interactions among TLS \cite{Phillips87,faoro_interacting_2015}, which cannot incorporate stochastic \textcolor{black}{and discrete} strong interactions \textcolor{black}{following a Poisson process}, such as those from fluctuators \cite{gao_semiempirical_2008,kumar_temperature_2008,faoro_internal_2012,Burn14,faoro_interacting_2015,de_graaf_quantifying_2021}. Fluctuators can be modeled as incoherent TLS whose $\Gamma_2^\text{ph} \ge \varepsilon$, as opposed to the coherent TLS introduced above in STM\cite{faoro_interacting_2015}. If strongly coupled with the coherent resonant TLS, the fluctuators can move the latter in and out of resonance with a jump rate $\gamma$ and effectively create a random telegraphic noise on the energy level $\varepsilon \rightarrow \varepsilon +\xi(t)$. \textcolor{black}{The fluctuators can be modeled as following a thermally activated tunneling process with rate $\gamma =\gamma_0 \exp(\frac{-E_\text{a}}{k_\text{B}T})$, where $E_\text{a}$ is the activation energy.  For a uniform distribution of $E_\text{a}\in[E_\text{a,min},\ E_\text{a,max}]$, the distribution of the fluctuator rates is thus $P(\gamma)\sim1/\gamma$ in an exponentially wide range $[\gamma_\text{min},\gamma_\text{max}]$, with $\gamma_\text{min}=\gamma|_{E_\text{a}=E_\text{a,max}}\sim \text{constant in }T$ and $\gamma_\text{max}=\gamma|_{E_\text{a}=E_\text{a,min}}\propto\exp(\frac{-E_\text{a,min}}{k_\text{B}T})$ \cite{faoro_internal_2012, faoro_interacting_2015}. The random telegraphic noise with a slow jump rate $\gamma$  happens infrequently during the measurement time, and thus cannot be averaged over to contribute to the spectral diffusion as in Eq. (\ref{Tds_eq}).} The exact solution to the Bloch equation will depend on the relationship between $\gamma, \Omega, \Gamma_1^\text{max}, \text{and}\  \Gamma_2$. $\Gamma_1^\text{max}$ is abbreviated to $\Gamma_1$ for clarity in the following discussion, which mainly focuses on the interaction between fluctuators and one resonant TLS. Thus, the distribution of values of $\Gamma_1$ for an ensemble of TLS is not invoked until the last step of integration to calculate the loss, and is not relevant to the fluctuators-induced effect. 

When the jump rate $\gamma$ is slow compared to the dynamics of the resonant TLS characterized by the rates $\Omega,\Gamma_1,\Gamma_2$, the stationary solution similar in form to Eq.(\ref{singleTLS}) can still be used with the substitution $\varepsilon \rightarrow \varepsilon +\xi$. After averaging over the distribution of the fluctuator jumps $\xi$, the response of a single TLS weakly coupled to low-$\gamma$ fluctuators is obtained (See Supplementary Material Section \ref{low_gamma_fluct}):
\begin{eqnarray}
\chi_\text{res}=m\frac{\omega-\varepsilon/\hbar-i(\Gamma_2+\Gamma_\text{f}/\sqrt{1+\kappa})}{(\Gamma_2\sqrt{1+\kappa}+\Gamma_\text{f})^2+(\omega-\varepsilon/\hbar)^2}
\label{singleTLS_slow}
\end{eqnarray}
which has the same form as Eq.(\ref{singleTLS}) but with the width of the Lorentzian widened by \textcolor{black}{ $\Gamma_\text{f} \propto T\lesssim \Gamma_2$ due to the weakly-coupled low-$\gamma$ fluctuators \cite{faoro_interacting_2015}.} For a continuous distribution  of TLS such as $P(\varepsilon,\Delta_0)$, the total internal loss is calculated by integrating Eq.(\ref{singleTLS_slow}) over the distribution function \textcolor{black}{ $P(\varepsilon,\Delta_0)$ and $P(\gamma)$ in the range $\gamma \in [\gamma_\text{min},\Gamma_1]$ .}

\begin{eqnarray}
\textcolor{black}{
    \delta_\text{TLS}
=\frac{1}{\epsilon_\text{r}\epsilon_0}\int\int\int\int P(\varepsilon,\Delta_0)P(\gamma)\left(\frac{\Delta_0 d_0}{\varepsilon}\right)^2\frac{\cos^2{\theta}}{\hbar}  
 \frac{m}{\sqrt{1+\kappa}}\frac{(\Gamma_2\sqrt{1+\kappa}+\Gamma_\text{f})}{(\Gamma_2\sqrt{1+\kappa}+\Gamma_\text{f})^2+(\varepsilon/\hbar-\omega)^2}d\varepsilon \text{d}\Delta_0 \text{d}\theta \text{d}\gamma}
 \label{loss_TLS_slow}
\end{eqnarray}

Clearly, the last fraction in the integral is a Lorentzian which evaluates to a constant after integration over $\varepsilon$, resulting in the same prediction for internal loss as the STM \cite{faoro_interacting_2015}.  

On the other hand, when the dynamics of the resonant TLS is dominated by a fast jump rate, $\gamma \gtrsim \Omega,\Gamma_1,\Gamma_2$, a probabilistic description of the resonant TLS must be adopted. Instead of directly solving the linearized Bloch equations Eq.(\ref{s+bloch_sup},\ref{szbloch_sup}), the master equation or the evolution equation of the probability distribution $\rho (\vec S)$ of the Bloch vector $\vec S =(\langle S_x^1 \rangle,\langle S_y^1 \rangle,\langle S_z^0 \rangle)$ is introduced: \cite{faoro_internal_2012}, 
\begin{eqnarray}
\frac{\partial \rho}{\partial t}+\frac{\text{d}}{\text{d} \vec S}(\frac{\text{d} \vec S}{\text{d} t}\rho)=\gamma[\delta(S_z^0-m)\delta(S_x^1)\delta(S_y^1)-\rho]
\label{evol_eq}
\end{eqnarray}
where $\text{d}\vec S/\text{d}t$ is given in Eq.(\ref{s+bloch_sup}, \ref{szbloch_sup}) with a time independent $\varepsilon$ where the random jumps $\xi(t)$ are dropped since the fast jumps are averaged out over a long time. (See Supplementary Material Section \ref{hi_gamma_fluct}).  The TLS loss is then extracted by solving for the average $y$ component of $\vec S$, $\overline{\langle S_y^1 \rangle }= \int \rho \langle S_y^1 \rangle \text{d}\vec S$, and integrating the solution over the probability distribution of fast fluctuator jump rates $P(\gamma) \propto 1/\gamma$ where $\gamma \in [\text{max}(\Omega,\Gamma_2), \gamma_\text{max}]$, and the distribution of resonant TLS energies $P(\varepsilon,\Delta_0)$. The loss has a logarithmic dependence on power :

\begin{eqnarray}
\delta  = m \delta_0  \text{arcsinh}\left(\frac{\gamma}{\Omega}\right)\Biggr|_{\text{max}(\Omega,\Gamma_2)}^{\gamma_{\text{max}}} \xrightarrow{\gamma_\text{max}\gg\Omega \gg \Gamma_1, \Gamma_2} m \delta_0 \ln (\frac{\gamma_\text{max}}{\Omega})
\label{high_gamma_log}
\end{eqnarray}
This expression explains the high power limit of the data in Fig. \ref{Qi_T} where the losses from different temperatures converge to a linear trend in the linear-log plot. This high $\gamma$ fluctuator loss will saturate to a constant value $\sim m\ln(\gamma_\text{max}/\Gamma_2)$ once $\Omega \lesssim \Gamma_2$ (See Supplementary Material Section \ref{hi_gamma_fluct}). Thus, it will not affect the low power behavior of the TLS loss.

More complicated is the case of intermediate jump rates where $\gamma \sim \Omega,\Gamma_1, \Gamma_2$. A similar master equation as in Eq.(\ref{evol_eq}) needs to be solved with $\varepsilon \rightarrow \varepsilon + \xi_k$ for each different fluctuator state $k$. The jumps are no longer ignored since their rates are close to the other dynamics ($\Omega, \Gamma_2, \Gamma_1$) in the system. After obtaining the average solution $\overline{\langle S_y^1 \rangle }$ for the TLS with energy levels $\varepsilon_k$, the same recipe for the loss calculation can be applied, namely, integrating the average solution over $P(\gamma)$ and then integrating over the distribution of the coherent TLS $P(\varepsilon,\Delta_0)$ . The loss is then ( See Supplementary Material Section \ref{im_gamma_fluct})

\begin{eqnarray}
    \delta=\delta_0 \int \frac{ m\Gamma_2}{\Gamma_2^2+(\varepsilon-\omega)^2}\left(
    \frac{1}{1+\kappa(\varepsilon)}\ln{\frac{\gamma_\text{h}}{\gamma_\text{l}}}+\frac{\kappa(\varepsilon)(1-n)}{(1+\kappa(\varepsilon))(1+n\kappa(\varepsilon))}\ln{\frac{1+\kappa(\varepsilon)+\gamma_\text{h}/\Gamma_1(1+n\kappa(\varepsilon))}{1+\kappa(\varepsilon)+\gamma_\text{l}/\Gamma_1(1+n\kappa(\varepsilon))}}
    \right) d\varepsilon
    \label{loss_imf}
\end{eqnarray}

where $\kappa(\varepsilon) = \Xi/\Gamma_1=\Omega^2\Gamma_2/[(\varepsilon/\hbar-\omega)^2+\Gamma_2^2]/\Gamma_1$, and $n$ is the probability that a given TLS is resonant, and is typically small for a system of many ($\sim10$) fluctuators (see supplementary material for \cite{faoro_internal_2012}), and thus ignored in the final model. $\gamma_\text{h,l}$ are the upper and lower bounds of the jump rates and are defined such that $\gamma_\text{h} \gtrsim \Xi+\Gamma_1,\sqrt{\Gamma_2^2+(\varepsilon/\hbar-\omega)} \gtrsim \gamma_\text{l}$. These limits translate to a range for power $\Omega$ where the power dependence of the loss is dominated by this model: $\Gamma_2  \gtrsim \Omega \gtrsim \sqrt{\Gamma_1\Gamma_2}$. Within this range, the loss from intermediate $\gamma$  fluctuators is approximately $\delta_0\ln[(\Omega^2+\Gamma_2^2)/(2\Omega^2)]$, a faster logarithmic power dependence than Eq. (\ref{high_gamma_log}). At higher  powers, the loss becomes constant $\sim m \ln(2)$. At lower power, the loss saturates to another constant $m \ln(\Gamma_2/\Gamma_1)$.\\
In summary, the three different fluctuation rates correspond to three different power ranges for the power dependence of the loss. In the high power limit  $\Omega \gtrsim \Gamma_2$, the effect of fluctuators that induce large $\gamma$ dominates and leads to a logarithmic power dependence; in the intermediate power regime $\Gamma_2 \gtrsim \Omega \gtrsim \sqrt{\Gamma_1\Gamma_2}$, the fluctuators with intermediate $\gamma$ give rise to a faster logarithmic power dependence, but meanwhile the saturation of TLS just as in STM has a comparable or even stronger power dependence and overlap in the same power regime; and finally in the low power limit $\Omega < \sqrt{\Gamma_1\Gamma_2}$, the typical TLS saturation in STM is recovered as the contributions from all three different types of fluctuators become constant in power. The above description qualitatively matches our experimental observation in Fig. \ref{Qi_T}.

\subsubsection{Fit to the internal loss measurements}
\begin{figure}[hbt!]
    \centering
    \includegraphics[scale=0.4]{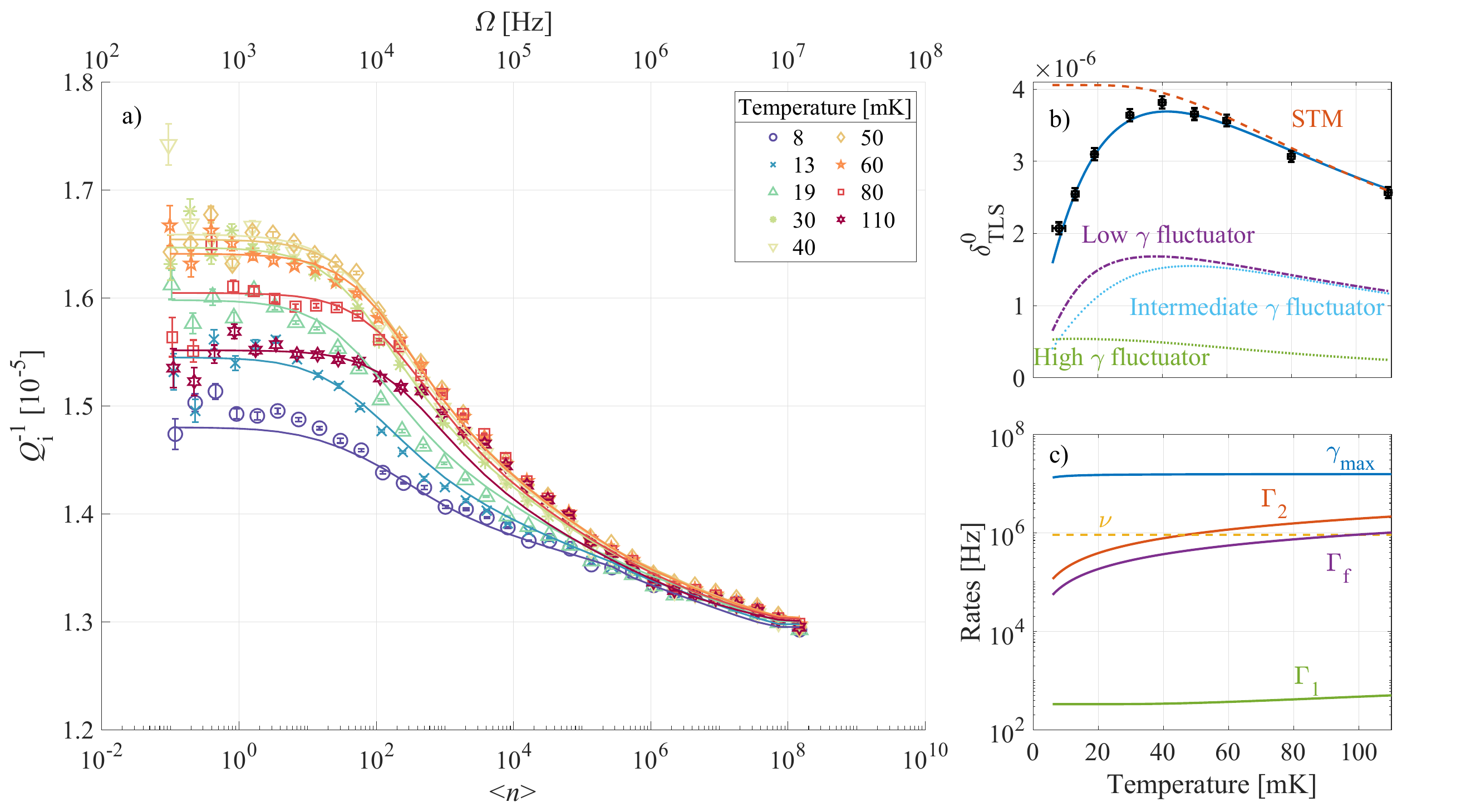}
    \caption{\label{fig:delta0_pc}  a) The least squares fit of the discrete GTM, \textcolor{black}{together with a constant background loss,} to the full power and temperature dependence of the measured internal loss below 150 mK b) Plot of $\delta_\text{TLS}^0(T)$ extracted from the average of the low power loss below TLS saturation in Fig. \ref{Qi_T}.  The orange dashed curve is the temperature dependence of STM loss below saturation power $\propto\tanh{(\varepsilon/(2k_\text{B} T))}$. The purple dash-dotted (light blue densely dotted) curve is from the discrete summation of individual TLS contributions for low (intermediate)-$\gamma$ fluctuators at zero applied power. The green dotted curve is the temperature dependent low power limit of the TLS loss induced by high $\gamma$ fluctuators. The blue solid curve is the sum of contributions from the low, intermediate, and high $\gamma$ fluctuators. c) Comparison of the temperature dependent rates determined from the least squares fit.}
\end{figure}

Although the power dependence of our data as in Fig. \ref{Qi_T} agrees with the effect of fluctuators in the GTM, the original model does not reproduce the observed temperature dependence. The GTM predicts the same temperature dependence of the TLS loss in the low power limit as in STM \cite{faoro_interacting_2015} shown as the orange  dashed curve in  Fig. \ref{fig:delta0_pc} (b), which clearly deviates from the extracted low power loss of TLS. To reconcile this difference, we propose a simple modification to the TLS model to account for the discrete coherent TLS near the resonance.  Consider the discrete form of the integral in the TLS loss for low $\gamma$ fluctuators, Eq. (\ref{loss_TLS_slow}),
%%%%%%%%%%%%%%%%%%%%%%%%%%%%%%%%%%%
\begin{comment}
To account for the smaller slope of $\delta_\text{TLS}(P_\text{ab})$ after TLS saturation, it is usually convenient to fit the power dependence of TLS loss with a free exponent $\beta$ smaller than 1/2 as in \cite{Macha10}:

\begin{equation}
    \delta_\text{TLS}=\delta_\text{TLS}^0(T)\left(1+\frac{P_\text{ab}}{P_\text{ab}C\text{c}(T)}\right)^{-\beta}+\delta_\text{other}
    \label{powerlaw_fit}
\end{equation}
where we emphasize the power dependence by lumping all the temperature dependent terms in the fitting parameters $\delta_\text{TLS}^0 (T)$ the zero power or intrinsic loss, and $P_\text{ab}C\text{c}(T)$ the critical power.  Here $\delta_\text{other}$ is the residual loss from all the other temperature and power independent loss mechanisms at low temperatures, such as the trapped vortices and the radiation loss dependent on the resonator geometry. Substituting $\beta=0.5$ in Eq. (\ref{powerlaw_fit}) recovers the STM formula in Eq. (\ref{STMloss}).\textcolor{black}{This equation is now commented out}

From the fit shown in Fig. \ref{Qi_T} using Eq. (\ref{powerlaw_fit}), two temperature dependent fitting parameters, $\delta_\text{TLS}^0 (T)$ and $P_\text{ab}C\text{c}(T)$, can be extracted and are plotted in Fig. \ref{fig:delta0_pc}.

\end{comment}
%%%%%%%%%%%%%%%%%%%%%%%%%%%%%%%%%%%%%%%%%%%

\textcolor{black}{
\begin{equation}
\delta_\text{TLS}=\frac{P_0d_0^2\Delta\varepsilon}{3\hbar\epsilon_\text{r}\epsilon_0}\ln\left(\frac{\Gamma_1}{\gamma_\text{min}}\right)\sum_n  \tanh(\frac{\varepsilon_n}{2k_\text{B} T}) \frac{\Gamma_2+\Gamma_\text{f}/\sqrt{1+\kappa}}{(\Gamma_2\sqrt{1+\kappa}+\Gamma_\text{f})^2+(\varepsilon_n/\hbar-\omega)^2} 
    \label{discrete_sum}
\end{equation}
}
where the index $n$ denotes the coherent TLS near the resonance and $\Delta\varepsilon$ is the average energy spacing in the TLS spectrum. We believe that Eq.(\ref{discrete_sum}) is justified since the number of coherent TLS inside the resonator bandwidth is $\sim 1$ for a TLS volume around $100 \mu m^3$ \cite{sarabi_cavity_2014}, and many previous works have observed individual TLS in microwave resonators \cite{lisen10,Grab12,Sarabi16_supporting,Chih22}. For the TLS exactly on resonance, $\varepsilon=\hbar\omega$, its loss \textcolor{black}{$\delta_\text{TLS}\sim \Gamma_2^{-1}\propto T^{-1}$} at low power, and is the classic result for the single TLS model in STM\cite{bhattacharya_jaynes-cummings_2011}. However, this stands in clear contrast to the observed reduction in loss at low temperature in Fig. \ref{fig:f0_Qi} (b) and Fig. \ref{fig:delta0_pc}. 

\textcolor{black}{It is thus required that the TLS is not always} on resonance ($\nu=\varepsilon_0/\hbar-\omega \neq 0$ where $\varepsilon_0$ stands for the energy level of the coherent TLS closest to resonance), a reasonable assumption given the sparse TLS distribution in the frequency spectrum for a small volume of TLS-inhabiting dielectrics. Mathematically, the width of the Lorentzian in the summation $w=\Gamma_2\sqrt{1+\kappa}+\Gamma_\text{f}$ dictates the transition from the low temperature reduced loss to the high temperature equilibrium result. For small $w$, a discrete sum will deviate from the integral since the Lorentzian is under-sampled. While for a Lorentzian with large $w$, a discrete sum with the same sampling rate will approximate the integral better. Specifically, at low powers ($\kappa \ll 1$), $w=\Gamma_2+\Gamma_\text{f}$  increases with the temperature and $w=\Gamma_2+\Gamma_\text{f} \sim \nu$ marks the transition temperature between the two regimes. For low temperatures ($w\ll\nu$), the Lorentzian term becomes roughly proportional to $w=\Gamma_2+\Gamma_\text{f}$ which gives the almost linear temperature dependence of loss. For higher power, $w$ increases with $\kappa$, which pushes the transition temperature lower and suppresses the low temperature reduction in loss. And eventually at high powers ($\kappa \gg 1$) such that $w > \nu$ for all temperatures, the equilibrium temperature dependence $m=(1/2)\tanh{(\hbar \omega /(2k_\text{B} T))}$ in STM is recovered in the entire temperature range. The same discrete summation can be applied to Eq. (\ref{loss_imf}) for intermediate $\gamma$ fluctuators. On the other hand,  Eq. (\ref{high_gamma_log}) for high $\gamma$ fluctuators is only modified with the substitution $\Gamma_2\rightarrow \sqrt{\Gamma_2^2+\nu^2}$ due to the sparse TLS assumption (See Supplementary Material Section \ref{hi_gamma_fluct}).  The final model that combines all three contributions is able to reproduce the full temperature ($T=8-110$ mK) and power ($\langle n\rangle =10^{-1}-10^8$) dependence of the loss shown as the solid curves in Fig. \ref{fig:delta0_pc} (a).

\textcolor{black}{The fit shows reasonable agreement with the data, with root mean squared error $\text{RMSE}=0.0124$. There are in total 10 fitting parameters, fewer degrees of freedom compared to fitting the data from different temperatures individually. The different contributions to the loss below TLS saturation power are plotted in Fig. \ref{fig:delta0_pc} (b) illustrating that the discrete TLS coupled to low and intermediate $\gamma$ fluctuators are responsible for the loss reduction. The different rates in the model determined from the fit are summarized in Fig. \ref{fig:delta0_pc} (c). The numerical values for $\Gamma_{1,2}$ and $\gamma_\text{max,min}$ are typical for TLS in amorphous materials \cite{faoro_interacting_2015}.  The rates also satisfy the following assumptions in the model: $\Gamma_2 \gtrsim \Gamma_\text{f}$, and  $\gamma_\text{max} \gg \Gamma_2$. In addition, the low temperature loss reduction occurs around $40$ mK as expected, when $\Gamma_2+\Gamma_\text{f} < \nu$, the width of the response is smaller than the detuning between TLS and the resonance. The other quantities extracted from the fit are listed below: the volume of TLS-inhabiting dielectrics, $10\ \mu  m^3$, the intrinsic TLS loss, $\delta^{TLS}_0= 3.85\times10^{-6}$, the other loss, $\delta_\text{other}=1.29\times10^{-5}$, and the minimum fluctuator rate $\gamma_\text{min}=4.5\times10^{-2} \text{ Hz}$.}
%%%%%%%%%%%%%%%%%%%%%%%%%%%%%%
\begin{comment}
Although Eq.(\ref{discrete_sum}) is derived from STM, it can be generalized to GTM in the limit of large $\gamma$ by replacing $\Gamma_1,\Gamma_2$ with $\gamma$. The same results would be obtained if we assume a linear temperature dependence of $\gamma \propto T$ similar to $\Gamma_2$. 

\textcolor{black}{What about discussion of $P_\text{ab}C\text{c}$?}

A Jaynes-Cummings model of a single TLS strongly coupled to a microwave photon predicts $\delta_\text{TLS}=\frac{1}{3\omega}(\Gamma_1+\Gamma_2)$ \cite{bhattacharya_jaynes-cummings_2011}, which would result in a linear reduction of loss, similar to our observation at low temperatures, and is shown as the orange dashed line in Fig. \ref{fig:delta0_pc}.
\end{comment}
%%%%%%%%%%%%%%%%%%%%%%%%%%%%%%

\section{Discussion}
The discrete and detuned TLS formalism will not affect the high $\gamma$-fluctuator contribution to internal loss, since the width of Lorentzian in the calculation of loss of high $\gamma$ fluctuators $w$ is widened by $\gamma$ such that $w\sim \gamma > \Delta\varepsilon/\hbar$ (See Supplementary Material Section \ref{hi_gamma_fluct}), which is indicated by the almost flat region in the green dotted curve at low temperature in Fig. \ref{fig:delta0_pc} (b).  However, the loss from intermediate $\gamma$ fluctuators could be subject to the low coherent TLS density but to a lesser degree than that from the low $\gamma$ fluctuators, since although the bandwidth of their response $\sim \Gamma_2$ is the same  (See Supplementary Material Section \ref{im_gamma_fluct}), there are many intermediate-$\gamma$-fluctuator-induced sublevels for one TLS in one Rabi cycle which effectively increases the density of available TLS energy levels. In order to avoid over fitting, this effect was not included in the model where the same density of states for TLS are assumed for those coupled to intermediate $\gamma$ fluctuators and the low $\gamma$ fluctuators.  Thus, the same $\Delta\varepsilon$ value is shared for the two different contributions. This simplification could lead to an underestimation of the loss in the intermediate power region, as illustrated by the deviation between the fit and data from $\langle n\rangle =10^2 \text{ to } 10^6$. 

The discrete TLS formalism only approximates the effect of a sparse TLS spectral density where despite the spectral diffusion with a width $\Gamma_2$, and the random telegraph noise characterized by the rate $\gamma$, the coherent TLS spends most of its time detuned from the resonance. The assumptions of even energy spacing  between TLS, $\Delta\varepsilon$, and constant energy levels, are convenient for numerical evaluation of the model, but are not necessary to reproduce the loss reduction at low temperature. Two other estimations of the probability of the TLS being on resonance, as well as the number of strongly coupled fluctuators that can bring a detuned TLS into resonance, are given in Supplementary Material Section \ref{low_gamma_fluct}. Both calculations show that for any TLS with a spectral width $\Gamma_2$ and a detuning to the resonance $\nu$, the TLS becomes less likely to be on resonance once $\Gamma_2(T)<\nu$ with  decreasing temperature, qualitatively agreeing with the experimentally observed loss reduction at low temperature.

\begin{comment}
However, this reduced loss would not affect our low power fit since the intermediate fluctuators contribution is weak and the discrete TLS will only further decrease the low power limit of the loss. The discrete and detuned TLS will only affect the range of significant power dependence for intermediate $\gamma$ fluctuators.
\end{comment}

The treatment above is largely classical where the TLS are treated as dipoles under classical field. A quantum mechanical approach that studies the Jaynes-Cummings model of a single TLS strongly coupled to a photon predicts a linear temperature dependence of the loss similar to our observation \cite{bhattacharya_jaynes-cummings_2011}. However, it should be noted that the photon frequency in our measurement (3.64 GHz) corresponds to weak photon-TLS coupling, since the Rabi frequency from the effective field of a single photon is much weaker than the relaxation rates $\Gamma_{1,2}$.  Additionally, the loss from strongly coupled TLS is predicted to show saturation in power at $\langle n \rangle \sim 1$, clearly lower than the observed saturation in the data at $\langle n \rangle \sim 10$ which corresponds to the weak coupling regime and reproduces the classical result \cite{bhattacharya_jaynes-cummings_2011}.

Although the fluctuators significantly affect the TLS internal loss, they should have limited effect on the frequency shifts \cite{faoro_interacting_2015}. The proposed discrete and detuned TLS formalism would not modify the STM frequency shift prediction either, because unlike the Lorentzian response function that governs the internal loss, the response function for frequency shift does not have a resonant shape and is not sensitive to the reduced sampling from the discrete TLS.

Ever since the importance of TLS interactions in amorphous solids was recognized by Yu and Leggett \cite{YuLeg88_supporting}, there have been numerous experimental works demonstrating evidence of TLS interactions,\cite{gao_semiempirical_2008,Burn14,lis15} and theoretical works treating the interacting TLS beyond STM \cite{coppersmith_frustrated_1991,burin_low_1994,vural_universal_2011}, with a recent example by Burin and Maksymov where they used a similar Master equation formalism \cite{Burin18}. However, \textcolor{black}{the fluctuations in the energy levels are averaged over to form the spectral diffusion}, unlike the fluctuators introduced by Faoro and Ioffe \cite{faoro_internal_2012,faoro_interacting_2015}, and the loss is predicted to have a power dependence faster than STM by a logarithmic factor, contrary to our observation.

At higher temperatures (above 150 mK), the quasiparticle effects become important, which corresponds to the upturn in loss in Fig. \ref{fig:f0_Qi}. The quasiparticles loss is related to its density $n_\text{qp}$ as,
\begin{align}
\frac{1}{Q_\text{qp}}&=\frac{2\alpha}{\pi}\frac{\sinh(\zeta)K_0(\zeta)n_\text{qp}}{N_0\sqrt{2\pi k_\text{B}T\Delta_{\text{S}0}}} \nonumber\\
\text{where}\quad
n_\text{qp}&=n_\text{th}+n_\text{noneq}=2 N_0 \sqrt{ 2 \pi k_\text{B} T \Delta_{\text{S}0} }  
\exp \left(-\frac{\Delta_{\text{S}0}}{k_\text{B} T } \right)+n_\text{noneq} 
\label{n_qp_analyticEq}
\end{align}
where $n_\text{noneq}$ is the non-equilibrium quasiparticle density. Similar to the fit for frequency shift, the model with only $n_\text{th}$ matches our data with the same set of fitting parameters $\Delta_0=170\ \mu eV$ and $\alpha =0.014$. 
%%%%%%%%%%%%%%%%%%%%%%%%%%%%%%%%%
\begin{comment}
 In addition, there are some temperature independent loss mechanisms including moving vortices arising from stray magnetic field and geometry dependent microwave radiation loss, etc. The identity $\frac{1}{Q_\text{others}} = \frac{1}{Q_{vortices}} +\frac{1}{Q_\text{rad}} $ can be used to summarize these temperature independent loss mechanisms. 

Therefore, one can say the internal quality factor of this half wavelength resonator is given by the following expression in the limit $T\ll T_\text{c}$: 
\begin{eqnarray}
\frac{1}{Q_\text{i}} =\frac{1}{Q_\text{TLS} (T,E)}+\frac{1}{Q_\text{qp} (T,E)}+ \frac{1}{Q_\text{others}} 
\label{One_Over_Qi}
\end{eqnarray}
\end{comment}
%%%%%%%%%%%%%%%%%%%%%%%%%%%%%%%%
 A calculation of the increased quasiparticle density including both thermal and non-equilibrium quasiparticles at high photon numbers in the half wavelength resonator based on Mattis-Bardeen equations \cite{Mattis-Bardeen_supporting,Turneaure} can be found in Supplementary Material Section \ref{nonep_qp_sup}. However, the results lack any strong temperature or power dependence below 100 mK. Note that this calculation includes the dynamics of the non-equilibrium quasiparticle finite lifetime due to recombination and trapping, with and without photon illumination \cite{Rothwarf,Parker_supporting,Guruswamy,Grunhaupt}.

\begin{comment}

The upturn in $Q_\text{i}(T)$ at low temperatures is due to an increase in TLS coherence times $T_2 \sim 1/T$, in this model. Although there may be no cause-and-effect relationship, it is interesting to note that such an increase can come about due to interactions between the TLS at low temperatures\cite{Phillips87,Burn14,Yu21}.  Evidence for interaction between TLS is well established from low temperature experiments on glasses.\cite{Carr94,Carr20}  Those studies have been interpreted as evidence of growing clusters of correlated TLS upon decreasing temperature.\cite{Yu21}  The TLS interact with each other by means of exchanging phonons, and can also be tuned by means of applied strain.\cite{Grab12,Lis15}  Direct measurements of $T_1(T)$ and $T_2(T)$ of individual TLS show significant enhancement for $T<300$ $mK$,\cite{Lisen10} consistent with the upturn in $Q_\text{i}(T)$ observed here.
\end{comment}

\section{Conclusion}
We have designed and fabricated capacitively-coupled half wavelength superconducting aluminum microwave resonators with minimum critical dimension of $1\ \mu m$ in the center conducting line of the CPW. The temperature and power dependence  of the resonator $Q_\text{i}$ deviate from the classical standard tunneling model results.  At high applied powers, the internal loss shows logarithmic power dependence, a signature of the generalized tunneling model with fluctuators.  At powers below TLS saturation, the internal loss decreases from 50 mK down to the fridge base temperature. We attribute this behavior to the detuning between TLS and the resonance frequency in a discrete TLS ensemble. Upon cooling, the single TLS response bandwidth, proportional to $\Gamma_2\propto T^{1.3}$, decreases. When the bandwidth drops below the detuning between TLS and the resonance frequency defined by the CPW resonator, the resonant TLS response decreases and contributes less to the internal loss. The generalized tunneling model is revisited and modified with the discrete TLS formalism resulting in a comprehensive fit to the measured loss in the entire low temperature and low power range, with a reasonable set of parameters.
%The increased $T_2$ of the many interacting TLS systems in the aluminum half-wavelength resonators inherits the nature of TLS loss.

%\medskip
%\textbf{Supporting Information} \par 
%Supplemental Material is available from the Wiley Online Library or from the author.

% Acknowledgements
\medskip
\textbf{Acknowledgements} \par %delete if not applicable))
This work is funded by the US Department of Energy through Grant \# DESC0018788 (sample preparation and measurements), and the National Science Foundation through Grant \# NSF DMR-2004386.  We acknowledge use of facilities at the Maryland Quantum Materials Center and the Maryland NanoCenter. We thank Rangga Budoyo and Braden Larsen for assistance with the non-equilibrium superconductor simulation.  We thank Dr. Ben Palmer and Yizhou Huang at Laboratory for Physical Sciences for assistance with the thermal evaporator and helpful suggestions.  

\nocite{*}

% References
\medskip

% Use the following code if you wish to generate your bibliography with BibTeX;
% replace the string "MSP-template" below with the name(s) of
% the BibTeX data base(s) you want to use.
% The resulting bibliography-output (the content of the .bbl file)
% must be pasted back into this file before submission.
% Please also include your BibTeX data base file(s) in your submission
% so that we can re-run BibTeX if necessary.
%
\bibliographystyle{MSP}
\bibliography{TaiTLS_bib}

\pagebreak

\setcounter{figure}{0}
\setcounter{equation}{0}
\setcounter{section}{0}
\makeatletter
\renewcommand{\figurename}{Fig.}
\renewcommand{\thefigure}{S\arabic{figure}}   % This adds S to the beginning of each Figure name

\renewcommand{\tablename}{Table}
\renewcommand{\thetable}{S\Roman{table}}    % This adds S to the beginning of each Table name

\renewcommand{\theequation}{S\arabic{equation}}    % This adds S to the beginning of each equation

\renewcommand{\thesection}{S\arabic{section}}
\makeatother

\title{Supplemental Material for Anomalous Loss Reduction Below Two-Level System Saturation in Aluminum Superconducting Resonators}% Force line breaks with \\

\maketitle
 
\date{\today}
\begin{abstract}
The Supplemental Material discusses the experimental details and measurement setup, reviews the standard and generalized tunneling models of two-level systems, reviews the master equation formalism for the fluctuators, presents an overview of the nonequilibrium quasiparticle and phonon model used to describe the nonequilibrium quasiparticle loss in the resonator, \textcolor{black}{and includes the finite element simulation of the resonator to determine the electric field coupled to the TLS dipoles.} 
\end{abstract}
\normalsize

\section{Experimental Details}\label{support Info_I}
\begin{figure} [ht]
\centering
\includegraphics[width=0.5\textwidth]{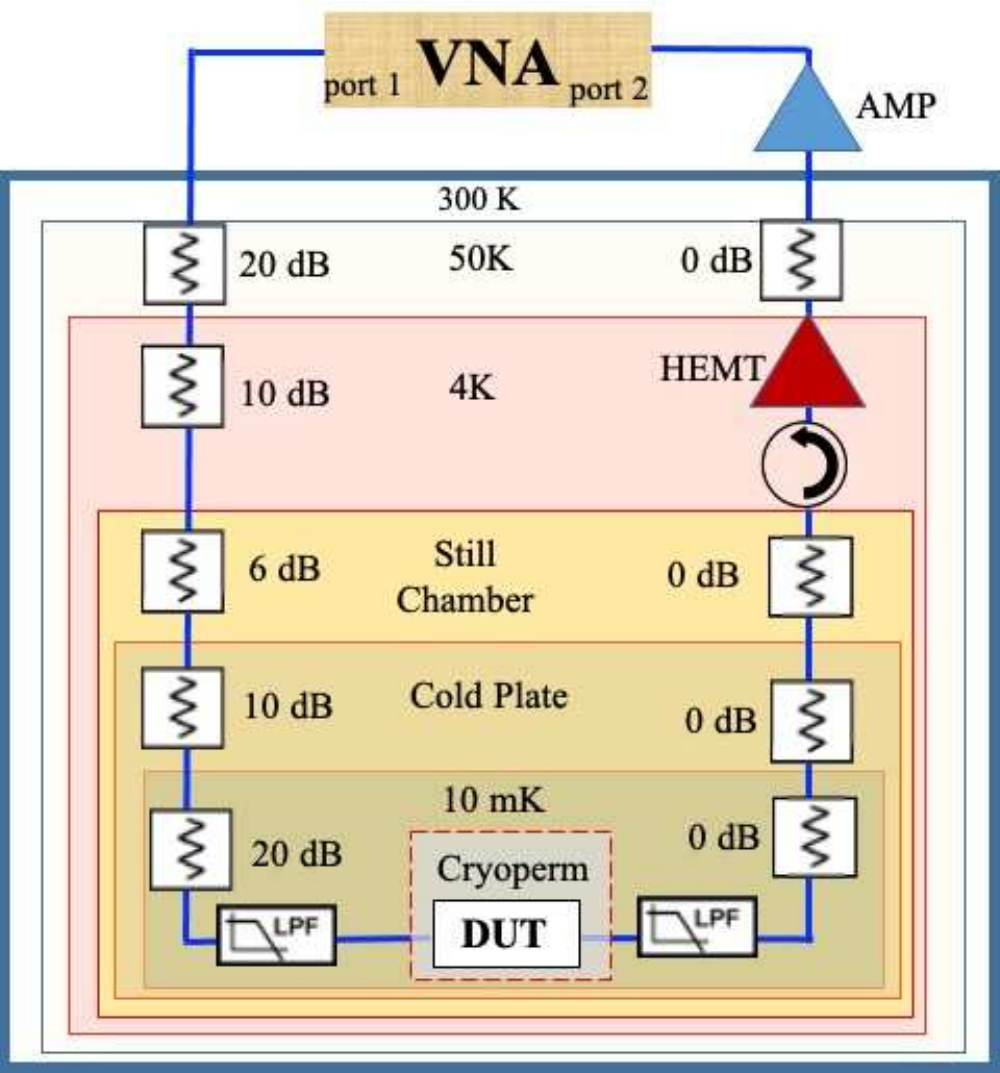}
\caption{\label{Setup}Schematic of the microwave measurement setup for the study of the aluminum $\lambda /2$ resonator.  The VNA at room temperature sends a signal from port 1 to the cryostat.  The signal is attenuated at each stage of the cryostat before passing through the low-pass filter (LPF) and entering the device under test (DUT).  The DUT is surrounded by a Cryoperm magnetic shield.  The output signal also passes through a low-pass filter before going through 0-dB attenuators that thermalize the coaxial cable center conductor.  The signal passes through an isolator and is amplified at the 4 K stage and at room temperature, before entering the VNA in port 2.}
\end{figure}

Fig. \ref{Setup} schematically shows the setup of the cryogenic transmission measurement with the VNA to characterize the resonator response. A variety of attenuators (produced by XMA) are used on each cryogenic stage to thermalize the center conductors of the coaxial cables.  The total attenuation in the input line is -66 dB.  Both the input line and output line on either side of the device are filtered by commercial microwave low-pass filters. The input line has a Marki Microwave low pass filter (FLP-1460) with 3-dB cutoff frequency at 14.6 GHz and the output line has another Marki Microwave low pass filter (FLP-1250) with 3-dB cutoff frequency at 12.5 GHz.  The output line goes through the cryogenic isolator (QUINSTAR Technology QCI-G0301201AM) with working frequency band 3-12 GHz, and then the signal is amplified by 36 dB using a commercial high-electron mobility transistor amplifier (Low Noise Factory \texttt{LNF-LNC0.3\_14A} with typical noise temperature 4.2 K) at the 4K stage, and then the transmitted signal is further amplified by 37 dB using another room temperature amplifier (HEMT) (Low Noise Factory \texttt{LNF-LNC2\_6A} with typical noise temperature 50 K at ambient temperature).

\pagebreak

\section{Conventional model for TLS loss}\label{STM_sup}

The TLS formalism is based on a simple model for a single TLS that can be described by the Hamiltonian,
$H_\text{TLS}=\frac{1}{2}
\begin{pmatrix}
-\Delta   &  \Delta_0 \\
 \Delta_0 &  \Delta 
\end{pmatrix}$
, where $\Delta $ is the asymmetry of the double well potential and $\Delta_0$ is the tunneling barrier energy between the  potential wells \cite{Phillips87,Budoyo2015}.  Here $\varepsilon=\sqrt{\Delta^2+\Delta_0^2}$ is the energy splitting of the two eigenvalues of $H_\text{TLS}$. A typical resonator hosts an ensemble of TLS with different values of $\Delta$ and $\Delta_0$. In the standard tunneling model (STM), a uniform distribution of the ensemble in $\Delta$ and log uniform distribution in $\Delta_0$ is assumed \cite{Phillips87}. Whereas in the generalized tunneling model (GTM), the distribution takes on an additional weak dependence of $\Delta$ due to a logarithmic correction \textcolor{black}{to the density of states for a system of very strongly interacting TLS \cite{cuevas_density_1989,faoro_interacting_2015,churkin_anomalous_2021}}. The distribution function can be mathematically expressed as

\begin{align}
    P(\Delta,\Delta_0)=(1+\mu)\left(\frac{\Delta}{\Delta_\text{max}}\right)^\mu\frac{P_0}{ \Delta_0}
    \label{distr_sup}
\quad   
\begin{cases}
    \mu=0 & \text{weakly interacting TLS}\\
    \mu\approx 0.3 & \text{strongly interacting TLS}
\end{cases} 
\end{align}
where $P_0\approx10^{44} J^{-1}m^{-3}$ is the density of states for TLS, $\Delta_\text{max} \sim \varepsilon_\text{max} \sim 100 K$ \cite{faoro_interacting_2015}, and the exponent $\mu$ approximately characterizes the logarithmic correction to the density of states of TLS due to strong TLS interactions \cite{cuevas_density_1989}.

The dynamics of a single TLS can be described by the linearized Bloch equations of the pseudospin $\vec S(t)=\vec S^0(t)+\vec S^1(t)$, where $\vec S^0$ is the solution to the homogeneous system without the external field, and $\vec S^1$ is the linear solution with frequency $\omega$ of the driving field. Under the rotating wave approximation, the Bloch equations become:  \cite{Phillips87,Gao_phdthesis,faoro_internal_2012} 
\begin{align}
i\frac{\text{d}\langle S^+ \rangle}{\text{d}t}=\Omega \langle S^0_z \rangle-(\omega-\varepsilon/\hbar+i\Gamma_2)\langle S^+ \rangle
\label{s+bloch_sup}
\\
\frac{\text{d}\langle S^0_z \rangle}{\text{d}t}=\Omega \Im \langle S^+\rangle -\Gamma_1\left( \langle S^0_z \rangle-m) \right)
\label{szbloch_sup}
\end{align}
with the Rabi frequency $\Omega$ characterizing the absorbed power, $\Omega \propto \sqrt{P_\text{ab}}$, and is defined as
\begin{equation}
\Omega= \frac{2 d_0 \Delta_0 }{\hbar \varepsilon} |\Vec{E}| 
\label{Rabi_Freq}
\end{equation}
where $S^+=S^1_x+iS^1_y$, $\Gamma_1$ and $\Gamma_2$ are the two phenomenological rates that describe the longitudinal ($S_z$) and transverse ($S_{x,y}$) relaxations, and $m=\tanh (\varepsilon/(2k_\text{B} T))/2$ is the equilibrium value of the $\langle S_z^0 \rangle$. In STM, the dielectric response of a single TLS can be obtained from the stationary solution to $\langle S^+ \rangle$ \cite{And72,Schick77,faoro_internal_2012}:
\begin{equation}
\langle S^ +\rangle=\frac{m\Omega(\omega-\varepsilon/\hbar-i\Gamma_2)}{(\omega-\varepsilon/\hbar)^2+\Gamma_2^2(1+\Omega^2\Gamma_1^{-1}\Gamma_2^{-1})}
\label{singleTLS_sup}
\end{equation}
The single TLS loss corresponds to the imaginary part of the response function in Eq.(\ref{singleTLS_sup}) which is in the form of a Lorentzian in $\varepsilon/\hbar$ centered at $\omega$ with a width:
\begin{align}
    w=\Gamma_2\sqrt{1+\kappa}
    \label{lorentzian_width_sup}
\end{align}
where $\kappa=\Omega^2\Gamma_1^{-1}\Gamma_2^{-1}$. The total dielectric loss is simply the integral of the single TLS contribution Eq.(\ref{singleTLS_sup}) over the distribution of the TLS Eq.(\ref{distr_sup}) \cite{Schick77,Phillips87,Gao_phdthesis}.
\begin{equation}
\delta_\text{TLS}
=\frac{1}{\epsilon_\text{r}\epsilon_0}\int\int\int P(\varepsilon,\Delta_0)\left(\frac{\Delta_0 d_0}{\varepsilon}\right)^2\frac{\cos^2{\theta}}{\hbar}  
 \frac{m\Gamma_2}{\Gamma_2^2(1+\kappa)+(\varepsilon/\hbar-\omega)^2}\text{d}\varepsilon \text{d}\Delta_0 \text{d}\theta
 \label{loss_int_sup}
\end{equation}
where $\epsilon_\text{r}\epsilon_0$ is the permittivity of the host dielectric material, $P(\varepsilon,\Delta_0)$ is obtained from Eq.(\ref{distr_sup}) with a change of variable from $\Delta$ to $\varepsilon$, $d_0$ is the maximum transition electric dipole moment of the TLS with energy splitting $\varepsilon$, $\vec{E}$ is the applied microwave electric field on the TLS dipole, and $\theta$ is the angle between the applied electric field and the TLS dipole moment.

For a typical TLS with $\varepsilon/h \approx$ 5 GHz and at reasonably low temperatures and powers, the width of its response $ w \approx\Gamma_2 \sim 1\ \text{MHz} \ll \omega$. Due to this sharp Lorentzian response function, the total loss is dominated by the resonant TLS whose energies $\varepsilon \sim \hbar\omega$. Before analytically evaluating the integral in Eq.(\ref{loss_int_sup}), the expressions for $\Gamma_{1,2}$ need to be introduced.

The longitudinal relaxation rate $(\Gamma_1)$ of a single TLS is dominated by the phonon process: \cite{Jack72,Hunk76,Bla77,Gao_phdthesis}
\begin{eqnarray}
\Gamma_1 = \left(\frac{\Delta_0}{\varepsilon}\right)^2
\left[\frac{\gamma_\text{L}^2}{v_\text{L}^5 }+\frac{2\gamma_\text{T}^2}{v_\text{T}^5 } \right] 
\frac{\varepsilon^3}{2 \pi \rho \hbar^4 } \coth (\frac{\varepsilon}{2 k_\text{B} T}) = \left(\frac{\Delta_0}{\varepsilon}\right)^2 \Gamma_1^\text{max}
\label{T1_eq_sup}
\end{eqnarray}
where  $\gamma_\text{L}$ and $\gamma_\text{T}$ are the longitudinal and transverse deformation potentials, respectively,  $v_\text{L}$ and $v_\text{T}$ are the longitudinal and transverse sound velocities, $\rho$ is the mass density, and $\Gamma_1^\text{max}$ is the maximum $\Gamma_1$ for the TLS with energy splitting $\varepsilon$, when $\Delta_0=\varepsilon$.

The transverse relaxation rate $(\Gamma_2)$ is defined as
\begin{align}
\Gamma_2=\Gamma_2^\text{ph}+\Gamma_\text{ds} \nonumber&\\
\text{where}\quad\Gamma_2^\text{ph}=\Gamma_1/2 \quad\text{and}\quad \Gamma_\text{ds}\sim & 10^{-3}(k_\text{B} T/\varepsilon_\text{max})^\mu k_\text{B} T/\hbar \quad\text{\cite{jankowiak_origin_1993,faoro_interacting_2015}}
\label{Tds_eq_sup}
\end{align}
$\Gamma_\text{ds}$ is the dephasing rate of the resonant TLS energy level $\varepsilon$, caused by its interactions with thermally activated TLS whose $\varepsilon \lesssim k_\text{B} T$, and is valid for low temperature measurement ($T<1$ K) \cite{jankowiak_origin_1993}.

After substituting the expressions for $\Gamma_1$, $\Gamma_2$ and $\Omega$ from Eqs. (\ref{T1_eq_sup}, \ref{Tds_eq_sup}, \ref{Rabi_Freq}) into the integral for the loss Eq.(\ref{loss_int_sup}) and using $\mu=0$, the famous STM prediction of TLS loss is obtained, \cite{Hunk76}
\begin{equation}
    \delta_\text{TLS}=\frac{\pi P_0d_0^2}{3\epsilon_\text{r}\epsilon_0}\frac{\tanh{\frac{\hbar\omega}{2k_\text{B} T}}}{\sqrt{1+(\Omega/\Omega_\text{c})^2}}
    \label{STMloss_sup}
\end{equation}
where $\Omega_\text{c} \propto \sqrt{\Gamma_1^\text{max}\Gamma_2}$ is the critical Rabi frequency that characterizes the saturation of TLS. The loss is expected to have an inverse square root dependence on power after the TLS saturation, $\delta_\text{TLS}\sim \Omega \propto P_\text{ab}^{-0.5}$ for $\Omega \gg \Omega_\text{c}$, which is much faster than observed in the data in Fig. \ref{Qi_T}.  This leads us to look at the generalized tunneling model with the inclusion of fluctuators.

\section{Master equation formalism for fluctuators}
\textcolor{black}{The theoretical work on master equation of the fluctuator coupling to TLS is based on \cite{faoro_internal_2012, faoro_interacting_2015}. The following discussion summarized their findings in a coherent manner and derives several key results based on the model. Much like the STM, the dynamics of a single TLS is described in the similar Bloch equations with an additional time dependent fluctuation of the TLS energy level $\xi(t)$ due to the fluctuators' interaction with the coherent TLS \cite{Phillips87,Gao_phdthesis,faoro_internal_2012}:}
\begin{align}
i\frac{\text{d}\langle S^+ \rangle}{\text{d}t}=\Omega \langle S^0_z \rangle-(\omega-(\varepsilon+\xi(t))/\hbar+i\Gamma_2)\langle S^+ \rangle
\nonumber
\\
\frac{\text{d}\langle S^0_z \rangle}{\text{d}t}=\Omega \Im \langle S^+\rangle -\Gamma_1\left( \langle S^0_z \rangle-m) \right)
\nonumber
\end{align}

The solution to the Bloch equations depends on the relation between the transition rates induced by fluctuators $\gamma$, \textcolor{black}{ whose distribution follows $P(\gamma)\sim 1/\gamma$ in an exponentially large range $[\gamma_\text{min},\gamma_\text{max}]$}, and the other inherent dynamics: $\Omega, \Gamma_{1,2}, \omega-\varepsilon/\hbar$. To systematically present the predictions from the different solutions to the mater equations under different limits, this section is structured according to the three regimes of the rate $\gamma$, namely high $\gamma > \Omega, \Gamma_2$, intermediate $\gamma \approx \Omega \text{ or } \Gamma_2$, and low $\gamma <\Gamma_1$. 

\subsection{Low \texorpdfstring{$\gamma$}{gamma} fluctuators} \label{low_gamma_fluct}
For low $\gamma$ fluctuators \textcolor{black}{($\gamma_\text{min}<\gamma<\Gamma_1$)}, the single TLS response is given by the stationary solution to Eq.(\ref{s+bloch_sup},\ref{szbloch_sup}), since the TLS dynamics is much faster than the jump rates $\gamma$. We can further separate the analysis for two different types: the common weakly-coupled fluctuators which shift the energy level of TLS by a  small energy $\xi$ and result in a widening of the spectral width of the TLS, and the rare strongly-coupled fluctuators whose $\xi$ is large enough and  produce large stochastic jumps on the TLS energy level $\varepsilon$. The weakly-coupled fluctuators induce detunings that follow a Lorentzian distribution with width $\Gamma_\text{f} \propto (k_\text{B} T/\hbar)(k_\text{B} T/\varepsilon_\text{max})^\mu \lesssim \Gamma_2 $ similar to the spectral diffusion\cite{faoro_interacting_2015}:
\begin{eqnarray}
P(\xi(t))=\frac{1}{\pi}\frac{\Gamma_\text{f}}{\Gamma_\text{f}^2+\xi(t)^2}
\label{fluct_distr}
\end{eqnarray}
 The single TLS imaginary response under a single weak fluctuator can be obtained from Eq.( \ref{singleTLS_sup})  by replacing $\varepsilon$ with $\varepsilon+\xi(t)$ \cite{faoro_interacting_2015}: 
 \begin{equation}
\text{Im} \langle S^ +\rangle=\frac{-m\Omega\Gamma_2}{(\omega-(\varepsilon+\xi(t))/\hbar)^2+\Gamma_2^2(1+\kappa)}
\label{fluct_single_TLS}
\end{equation}
where $\kappa=\Omega^2/(\Gamma_1\Gamma_2)$. Integrating Eq.(\ref{fluct_single_TLS}) over the distribution of detuning $P(\xi(t))$ \textcolor{black}{and the distribution of the fluctuator rate $P(\gamma)$}, one can obtain the single TLS loss:
\textcolor{black}{
\begin{equation}
    \delta(\varepsilon)\propto\ln\left(\frac{\Gamma_1}{\gamma_\text{min}}\right)\frac{m}{\sqrt{1+\kappa}}\frac{\Gamma_2\sqrt{1+\kappa}+\Gamma_\text{f}}{(\Gamma_2\sqrt{1+\kappa}+\Gamma_\text{f})^2+(\omega-\varepsilon/\hbar)^2}
    \label{slow_fluc_STM}
\end{equation}
}
The second fraction is in the form of a Lorentzian with width $\Gamma_2\sqrt{1+\kappa}+\Gamma_\text{f}$, and the first term is the same as the STM formula. If a continuous distribution function for TLS $P(\varepsilon,\Delta_0)$ is assumed, one can obtain the total internal loss by integrating  Eq.(\ref{slow_fluc_STM}) over $P(\varepsilon,\Delta_0)$, resulting in the same internal loss as in STM.

The distribution $P(\xi(t))$ may not apply to the strongly-coupled fluctuators which completely shift the TLS in and out of resonance, since there are typically only a few strong fluctuators in surface dielectrics \cite{faoro_interacting_2015}. They contribute to the imaginary part of a single TLS response as a random telegraph noise (i.e. $\xi(t)$ in Eq.(\ref{fluct_single_TLS}) ). The internal loss is still in the form of the STM loss Eq.(\ref{singleTLS_sup}), \textcolor{black}{but could be reduced by the strong fluctuators occasionally moving the coherent TLS in and out resonance}.

Three methods were used to estimate the average effect of the low $\gamma$ fluctuators on the internal loss. Intuitively, we could estimate the total loss by summing up the contribution from all the TLS near the resonance. The key assumption is the low density of the resonant TLS which justifies the discrete treatment. An estimation based on \cite{sarabi_cavity_2014} yields $\sim 1$ TLS in the bandwidth of the resonance for a volume of TLS-inhabiting dielectrics around 30 $\mu m^3$, which supports our assumption. The first method is thus a summation of all the single TLS loss in the form of Eq.(\ref{slow_fluc_STM}) near the resonance with $\varepsilon =...\hbar(\omega+\nu)-2\Delta\varepsilon,\hbar(\omega+\nu)-\Delta\varepsilon,\hbar(\omega+\nu),\hbar(\omega+\nu)+\Delta\varepsilon,\hbar(\omega+\nu)+2\Delta\varepsilon,... $ , and is given in Eq.(\ref{discrete_sum}). A finite detuning less than the average energy spacing in the TLS spectrum, $\hbar\nu<\Delta\varepsilon$, is included such that the TLS closest to resonance will not give a diverging response as $T\rightarrow0$, and instead contributes to the loss as $m\Gamma_2\Delta\varepsilon/(\hbar\nu^2)\propto T^{1+\mu}$ at low temperature. This detuning is a consequence of a sparse TLS distribution where $\Delta\varepsilon/\hbar\gtrsim \Gamma_2$. Thus, any given TLS is rarely on resonance due to the low density of states. We should emphasize that this treatment is independent from the master equation formalism for the fluctuators, and can be applied to STM with low TLS density where the use of  distribution function $P(\varepsilon,\Delta_0)$ is inappropriate.

The second method assumes that the detuning $\nu$ is not much larger than $\Gamma_2$ so that the distribution $P(\xi(t))$ for the weakly-coupled fluctuators can still be applied to the case $\xi/\hbar \sim \nu \lesssim \Gamma_2$. The probability of a TLS with a detuning $\nu$ to be on resonance under the influence of the low $\gamma$ fluctuators can be calculated by integrating $P(\xi(t))$:
\begin{equation}
    P_{\text{on resonance}}=\left.\frac{1}{\pi}\arctan(\frac{\xi}{\hbar\Gamma_\text{f}})\right\rvert_{\hbar(\nu-\Gamma_2)}^{\hbar(\nu+\Gamma_2)}
    \label{weak_fl}
\end{equation}
which also leads to a monotonic increasing temperature dependence below 50 mK as in the data.

The third method estimates the number of strongly-coupled fluctuators capable of moving the coherent TLS in and out of resonance. Since the energy drift caused by fluctuators is directly related to the interaction energy, which is dipole-dipole like, $U(r)\sim r^{-3}$, we could find a volume around the near-resonant TLS which hosts fluctuators that interact strongly enough. For $\Gamma_2 <\nu$, the TLS is detuned, and the criterion for being a strong fluctuator is that the induced energy shift can bring the TLS in resonance such that $\hbar(\nu-\Gamma_2)<U(r)<\hbar(\nu+\Gamma_2)$. This volume is a spherical shell with inner and outer radius defined by the bounds on the interaction energy $U(r)$. For $\Gamma_2 > \nu$, the TLS is already in resonant, and the strong fluctuators are those that move the TLS out of resonance, with $U(r)>\hbar(\Gamma_2-\nu)$. This volume is a sphere with radius defined by $U(r)=\hbar(\Gamma_2-\nu)$. The total number of strong fluctuators can then be estimated as:
\begin{equation}
    N_{\text{strong fluctuators}}=(1+\mu)P_0\left(\frac{k_\text{B} T}{\varepsilon_\text{max}}\right)^\mu V_{\text{strong fluctuators}} k_\text{B} T\propto 
    \quad
    \begin{cases}
    \frac{2\Gamma_2^2}{\nu^2-\Gamma_2^2} \quad (\Gamma_2<\nu)\\
    \frac{\Gamma_2}{\Gamma_2-\nu} \quad (\Gamma_2>\nu)
    \end{cases}
    \label{strong_fl}
\end{equation}
where the density of states for TLS from Eq.(\ref{distr_sup}) is used, and the energy range is set to the thermal energy $k_\text{B} T$. The two expressions Eq.(\ref{weak_fl},\ref{strong_fl}) are plotted below as a function of $\Gamma_2\propto T^{1+\mu}$. 

\begin{figure} [ht]
\centering
\includegraphics[width=0.6\textwidth]{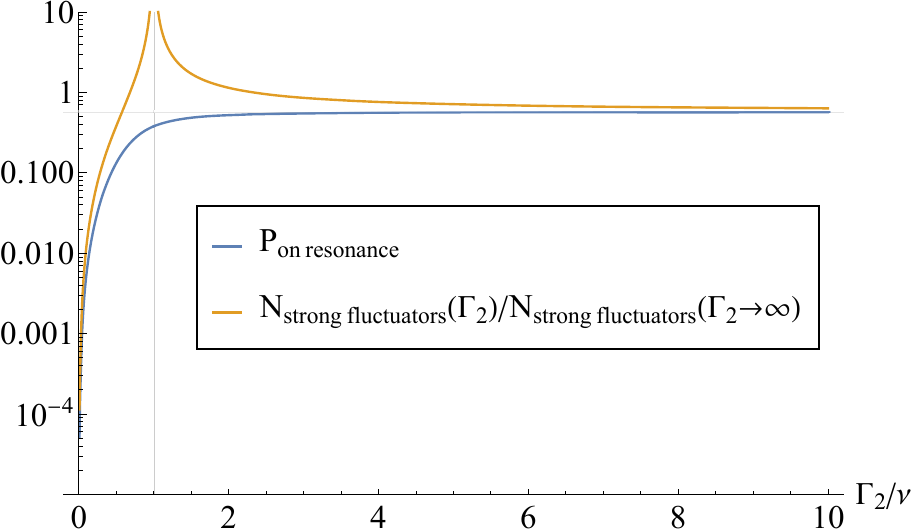}
\caption{\label{weak_strong_low_gamma} The blue curve is the probability for a TLS detuned with $\nu$ to be resonant under the influence of weakly-coupled fluctuators. The yellow curve is the normalized estimated number of strongly-coupled fluctuators. They are plotted as a function of $\Gamma_2\propto T^{1.3}$, which also represents their temperature dependence. It can be concluded that for both the strong and weak fluctuators, as temperature decreases ($\Gamma_2 /\nu \rightarrow 0$) the TLS are increasingly detuned.}
\end{figure}

The x axis of Fig. \ref{weak_strong_low_gamma}, $\Gamma_2$ is normalized by the detuning $\nu$. When $\Gamma_2$ increases to $\nu$, the probability for TLS to be resonant under the weak fluctuators (Blue curve in Fig. \ref{weak_strong_low_gamma}) approaches a constant $2/\pi\arctan(\Gamma_2/\Gamma_\text{f})$, which is denoted by the horizontal line where $\Gamma_\text{f} =0.8 \Gamma_2$. The interpretation of the number of strong fluctuators (Yellow curve in Fig. \ref{weak_strong_low_gamma}) is more nuanced. When $\Gamma_2 < \nu$, the coherent TLS itself is detuned, the strong fluctuators shift the TLS in resonance, and therefore their number is positively correlated with the TLS response. However, for $\Gamma_2 > \nu$, the strong fluctuators shift the already resonant TLS out of resonance, and thus their number is negatively correlated with the TLS response. Consequently, as $\Gamma_2$ or temperature increases, the resonant response becomes stronger. The divergence near $\Gamma_2 \sim \nu$ represents the failure of this estimation in the range where any arbitrarily small shift can move the TLS in and out of resonance, and the strong fluctuator volume goes to infinity. In reality, this volume is bounded by the host material for TLS. For $\Gamma_2 \gg \nu$, the case of TLS exactly on resonance is recovered, and the number of strong fluctuators is clearly independent of temperature. \textcolor{black}{A numerical estimate of $N_{\text{strong flucutators}}(\Gamma_2/\nu\rightarrow\infty)$ gives $\lesssim 1$ for typical weakly interacting TLS  \cite{faoro_interacting_2015}.} In both estimates, the detuned TLS becomes resonant more often as $\Gamma_2$ or temperature increases, which qualitatively explains the increase in loss in temperature at low temperatures in the measurement.

It should be noted that if applied to TLS exactly on resonance ($\Gamma_2/\nu\rightarrow\infty$), the above arguments will lead to the conclusion that low $\gamma$ fluctuators present no temperature dependence for both the strong and weak low $\gamma$ fluctuators, which forms the basis of the claim that no anomalous temperature dependence on the loss is expected in the original GTM work \cite{faoro_interacting_2015}.

\subsection{High \texorpdfstring{$\gamma$}{gamma} fluctuators}\label{hi_gamma_fluct}
For fluctuators that induce fast jump rates on the resonant TLS, $\gamma \gtrsim \Omega, \Gamma_{1,2}$,  their effects on the dynamics are described by the master equation \cite{faoro_internal_2012}:
\begin{eqnarray}
\frac{\partial \rho}{\partial t}+\frac{\text{d}}{\text{d} \vec S}(\frac{\text{d} \vec S}{\text{d}t}\rho)=\gamma[\delta(\langle S_z^0 \rangle-m)\delta(\langle S_x^1\rangle)\delta(\langle S_y^1 \rangle)-\rho]
\label{evol_eq_sup}
\end{eqnarray}
where $\vec S =(\langle S_x^1 \rangle,\langle S_y^1 \rangle,\langle S_z^0 \rangle)$ and $\rho$ is the probability for a TLS to be in the state $\vec S$. The case for high $\gamma$ fluctuators is particularly simple, where the fast random telegraph noise is averaged out. Thus, $\text{d} \vec S/\text{d}t$ is solved from Eq.(\ref{s+bloch_sup},\ref{szbloch_sup}) with $\xi(t) = 0$. The average y-component response of the single TLS with energy splitting $\varepsilon$ can then be solved from the master equation Eq.(\ref{evol_eq_sup}) as:
\begin{equation}
    \overline{\langle S_y^1 \rangle }= \int \rho \langle S_y^1 \rangle \text{d}\vec S=\frac{m\Omega}{\sqrt{1+\frac{\Omega^2}{(\gamma+\Gamma_1)(\gamma+\Gamma_2)}}}\frac{\sqrt{(\gamma+\Gamma_2)^2+\Omega^2\frac{\gamma+\Gamma_2}{\gamma+\Gamma_1}}}{(\gamma+\Gamma_2)^2+\Omega^2\frac{\gamma+\Gamma_2}{\gamma+\Gamma_1}+(\omega-\varepsilon/\hbar)^2}
    \label{higammasol}
\end{equation}
where again the second fraction is in the form of a Lorentzian with width $w$ given as
\begin{equation}
    w=\sqrt{(\gamma+\Gamma_2)^2+\Omega^2\frac{\gamma+\Gamma_2}{\gamma+\Gamma_1}}
\end{equation}
This width is guaranteed to be larger than $\nu$ for large $\gamma$. Therefore, we could still proceed with integrating the Lorentzian (the second fraction) in Eq.(\ref{higammasol}) over $\varepsilon$, which leads to a constant independent of temperature and power. After another integration over the distribution $P(\gamma)$, we can express the total internal loss as
\begin{equation}
    \delta = m \delta_0  \text{arcsinh}\left(\frac{\gamma}{\Omega}\right)\Biggr|_{\text{max}(\Omega,\sqrt{\Gamma_2^2+\nu^2})}^{\gamma_\text{max}}
    \label{higammaloss}
\end{equation}
The function $\text{arcsinh}(x)$ is reduced to $\ln(2x)$ for $x\gg1$ giving the observed logarithmic power dependence. And naturally for $\Omega \rightarrow 0$, $\delta\rightarrow m \delta_0 \ln(\gamma_{\text{max}}/\sqrt{\Gamma_2^2+\nu^2})$, which is a constant limit at low power much like the TLS saturation in STM.

The rate $\sqrt{\Gamma_2^2+\nu^2}$ is calculated from the characteristic rate of the TLS dynamics with $\xi(t)=0$ and characterizes the boundary between fast and intermediate jump rates $\gamma$. 

\subsection{Intermediate \texorpdfstring{$\gamma$}{gamma} fluctuators} \label{im_gamma_fluct}
When $\gamma$ is comparable to $\Omega$ and/or $\Gamma_{1,2}$, several jumps could be expected before the TLS relaxes to its ground state. Therefore, the interaction among the fluctuators need to be accounted for, which complicates the modelling compared to the low and high $\gamma$ fluctuators. The dynamical effect is still described by a similar master equation Eq.(\ref{evol_eq_sup}) for high $\gamma$ fluctuators with $\rho$ replaced by $\rho_k$ for each different state $k$ of the fluctuators. Moreover, the fast random telegraph noise is not averaged out, and the dynamics of the transverse component can be treated as stationary in a rotating wave approximation ($d\langle S^+ \rangle/dt=0$). The resulting master equation to be solved is then \cite{faoro_internal_2012}
\begin{align}
    &\frac{\partial \rho_k}{\partial t}+\frac{\text{d}}{\text{d}z}[(\Gamma_1(m-z)-\Xi_k z)\rho_k]=\gamma_{kn}\rho_n \label{im_master}\\
    &y_k=\frac{-\Omega\Gamma_2}{\Gamma_2^2+(\varepsilon_k/\hbar-\omega)^2}z_k
    \label{im_zy}
\end{align}
where $\langle S_z^0\rangle$, $\langle S_y^1\rangle$ are abbreviated as $z$ and $y$, and $\Xi_k=\Omega^2\Gamma_2/[(\varepsilon_k/\hbar-\omega)^2+\Gamma_2^2]$, and $\gamma_{kn}$ is the transition rate between the fluctuator state $k$ and $n$. $\Gamma_1$ and $m$ are not subscripted with $k$ since their $\varepsilon_k$ dependence contains a slowly varying function in the form $\tanh[\varepsilon_k/(2k_\text{B}T)]$ that should not vary much near the resonance. $z_k$ can be solved from Eq.(\ref{im_master}) and substituted in Eq.(\ref{im_zy}) to obtain $y_k$
\begin{align}
    y_k=\frac{-\Omega\Gamma_2}{\Gamma_2^2+(\varepsilon_k-\omega)^2}\frac{n_k}{1+\frac{1+n_k\gamma/\Gamma_1}{\gamma+\Gamma_1}\Xi_k}
    \label{im_single_sol}
\end{align}
where $n_k$ is the probability for a TLS to have energy $\varepsilon_k$, $\gamma=\sum_n (\gamma_{kn}+\gamma_\text{nk})$ is the total effective jump rate for fluctuator state $k$. The single TLS y-component response with the energy level $\varepsilon_k$ can then be calculated from integrating Eq.(\ref{im_single_sol}) over the probability distribution $P(\gamma)\propto 1/\gamma$
\begin{align}
      \overline{y_k}\propto \frac{-m\Omega\Gamma_2}{\Gamma_2^2+(\varepsilon_k-\omega)^2}\left(
    \frac{1}{1+\kappa_k}\ln{\frac{\gamma_\text{h}}{\gamma_\text{l}}}+\frac{\kappa_k(1-n_k)}{(1+\kappa_k)(1+n_k\kappa_k)}\ln{\frac{1+\kappa_k+\gamma_\text{h}/\Gamma_1(1+n_k\kappa_k)}{1+\kappa_k+\gamma_\text{l}/\Gamma_1(1+n_k\kappa_k)}}
    \right)
    \label{im_final_single}
\end{align}
where $\kappa_k =\Xi_k/\Gamma_1$ is reduced to $\kappa=\Omega^2/(\Gamma_1\Gamma_2)$ when the TLS is on resonance ($\varepsilon_k=\hbar\omega$). The upper and lower bounds for the intermediate fluctuators are obtained from the TLS dynamics as in Sec.\ref{hi_gamma_fluct} such that $\gamma_\text{h} \gtrsim \Xi_k+\Gamma_1,\sqrt{\Gamma_2^2+(\varepsilon_k/\hbar-\omega)^2} \gtrsim \gamma_\text{l}$. 

The total loss can then be obtained from summing up the contribution from all the different TLS energy levels just as in Eq.(\ref{discrete_sum}). 
\begin{eqnarray}
    \delta=\frac{\delta_0}{\pi}\sum_k\frac{m\Gamma_2}{\Gamma_2^2+(\varepsilon_k-\omega)^2}\left(\frac{1}{1+\kappa_k}\ln{\frac{\gamma_\text{h}}{\gamma_\text{l}}}+\frac{\kappa_k(1-n_k)}{(1+\kappa_k)(1+n_k\kappa_k)}\ln{\frac{1+\kappa_k+\gamma_\text{h}/\Gamma_1(1+n_k\kappa_k)}{1+\kappa_k+\gamma_\text{l}/\Gamma_1(1+n_k\kappa_k)}}\right)
    \label{discrete_sum_imf_sup}
\end{eqnarray}
If one assumes that the energy drift induced by the individual fluctuator follows the same Lorentzian distribution $P(\xi)$ for the weak fluctuators in Sec.\ref{low_gamma_fluct}, the total drift will follow a Lorentzian distribution with a wider width $P(\xi)\propto N\Gamma_\text{f}/((N\Gamma_\text{f})^2+\xi^2)$ since there are $N$ fluctuators affecting the TLS in the case of intermediate fluctuators. If  we assume that this width $N\Gamma_\text{f}$ is comparable to the detuning $\nu$ , then the broadening of the TLS energy levels will ensure the existence of resonant TLS. Therefore, one can apply the continuous distribution of TLS $P(\varepsilon,\Delta_0)$. The final result of the internal loss after the integration is then

\begin{align}
    \delta\propto \tanh (\frac{\hbar\omega}{2k_\text{B} T})\left(
    \frac{1}{1+\kappa}\ln{\frac{\gamma_\text{h}}{\gamma_\text{l}}}+\frac{\kappa(1-n)}{(1+\kappa)(1+n\kappa)}\ln{\frac{1+\kappa+\gamma_\text{h}/\Gamma_1(1+n\kappa)}{1+\kappa+\gamma_\text{l}/\Gamma_1(1+n\kappa)}}
    \right)
    \label{im_loss}
\end{align}

The loss takes on a logarithmic power dependence in the intermediate power range $\left( \sqrt{\Gamma_1\Gamma_2}<\Omega<\sqrt{\Gamma_2^2+\nu^2}\right)$,  $\delta\sim m\ln((\Omega^2+\Gamma_2^2)/(2\Omega^2))$. This loss saturates to a constant $\sim m\ln(\sqrt{\Gamma_2^2+\nu^2}/\Gamma_1)$ for lower powers, and to a small constant $\sim m \ln(2)$ for higher powers.
\begin{comment}
\begin{figure}[hbt!]
\includegraphics[width=0.9\textwidth]{best_discrete_sum_full_pow_dep_new.pdf}
\caption{\label{best_full_fit} The least squares fit of GTM with the discrete summation to the observed power dependence at different temperatures. The dot-dashed and dashed vertical lines represent the cut-offs for the high-and-intermediate, and intermediate-and-low power limits. The vertical dotted lines represent the characteristic saturation power from STM, $\Omega_c \propto \sqrt{\Gamma_1\Gamma_2}$, which clearly deviate from the observed power dependence.}
\end{figure}
\end{comment}

In summary, the power dependence of the loss can be separated into three different regimes where in the high power limit $\Omega \gtrsim \Gamma_2$, the effect of fluctuators that induce large $\gamma$ dominates and leads to a logarithmic power dependence; in the intermediate power regime $\Gamma_2 \gtrsim \Omega \gtrsim \sqrt{\Gamma_1\Gamma_2}$, the fluctuators with intermediate $\gamma$ give rise to a faster logarithmic power dependence, but meanwhile the saturation of TLS just as in STM has a comparable if not stronger power dependence and overlap in the same power regime; and finally in the low power limit $\Omega < \sqrt{\Gamma_1\Gamma_2}$, the typical TLS saturation in STM is recovered as the fluctuator contributions from all three different regimes become constant in power. The above description matches our experimental observation in Fig. \ref{Qi_T}.

The final fit to the data can be obtained from summing up all three contributions from different regimes of fluctuators: the discrete summation for the low and intermediate $\gamma$ fluctuators, Eq.(\ref{discrete_sum}) and Eq.(\ref{discrete_sum_imf_sup}) respectively, and the logarithmic power dependent loss from the high $\gamma$ fluctuators Eq.(\ref{higammaloss}), and is shown in Fig. \ref{fig:delta0_pc}.

\begin{comment}
The large difference between the data and fits at the intermediate power range is likely the consequence of the assumption $N \Gamma_\text{f} \sim \nu$ which might not always be satisfied. If the detuning is larger than the width of an individual TLS response bandwidth, a comparable reduction in loss to the discrete summation for the low $\gamma$ fluctuators should be expected. Similarly, a temperature dependent model of $\gamma_\text{max}$: $\gamma_\text{max}=\gamma_0 \exp{-E_\text{a}/(k_\text{B}T)}$ with nonzero minimum thermal activation energy $E_\text{a}$ could also be included to improve the fits. However, we do not incorporate either mechanism to the fit to avoid introducing too many degrees of freedom to the model.
\end{comment} 
\section{Alternative models for fitting TLS loss}\label{other_model}
None of the existing TLS models shows a strong temperature dependence of loss below TLS saturation. \textcolor{black}{Hence alternative models need to be combined with the discrete TLS formalism to account for the observed temperature dependence of loss. Here we examine two models commonly adopted to explain the slow power dependence above TLS saturation power, as well as an alternative version of the discrete GTM fit used in the main text in which an energy dependent density of states is assumed with $\mu \ne 0$.}
 
\subsection{Two TLS model}\label{alt_two_tls}
\begin{figure}[hbt!]
    \centering
    \includegraphics[width=0.9\textwidth]{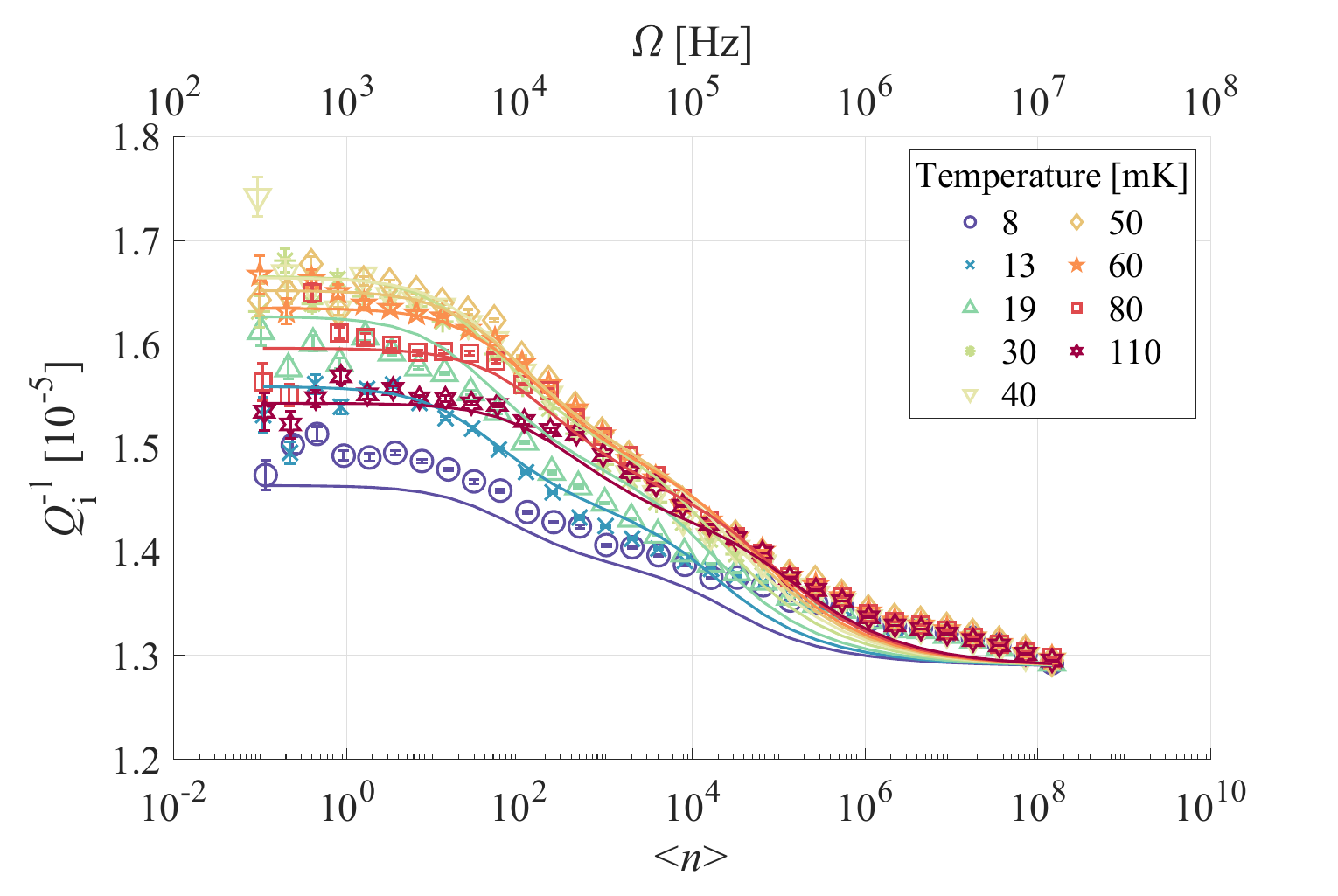}
    \caption{The discrete two TLS model fit to the loss data. }
    \label{fig:alt_two_tls}
\end{figure}

The two TLS or, by extension, multiple TLS model is a popular alternative in explaining the slower power dependence than STM. The multiple species of TLS could be attributed to nonuniform field distribution in the resonator \cite{khalil_2011}, different dipole moments of the TLS \cite{yu_experimentally_2022}, and different microscopic origins of the TLS \cite{schechter_inversion_2013}. The exact model for TLS loss is simply a sum of two different STM loss contributions:
\begin{eqnarray}
    \delta_\text{TLS} = \sum_n  \delta_0^{TLS,1} \frac{m_n\Gamma_2}{\Gamma_2^2(1+\kappa)+(\varepsilon_n/\hbar-\omega)^2}+\delta_0^{TLS,2} \frac{m_n\Gamma_2}{\Gamma_2^2(1+\eta \kappa)+(\varepsilon_n/\hbar-\omega)^2}   
    \label{alt_2tls_loss}
\end{eqnarray}
where $m_n=\tanh(\varepsilon_n/(2k_\text{B} T))$ and its subscript explicitly denotes its dependence on $\varepsilon_n$. The difference between the two species of the TLS is encoded in their intrinsic losses $\delta_0^{TLS,1},\delta_0^{TLS,2}$ which accounts for different participation ratios, and the factor $\eta$ which could be due to different field strength experienced by the TLS dipoles, different dipole moments, or different relaxation rates of the two species of TLS. The resulting fit in Fig. \ref{fig:alt_two_tls} shows a systematic deviation between the data and the fit, with $\text{RMSE}=0.021$.
There are 9 fitting parameters in this model. The resulting relaxation rates from the fit are $\Gamma_2=4.7\times10^5 \text{ to } 2\times10^7 Hz$ and $\Gamma_1=6.8\times10^3 \text{ to } 10^4 Hz$. The losses from the different contributions are determined as $\delta_\text{other} = 1.29 \times10^{-5}$, $\delta^{TLS,1}_0=6.9 \times10^{-6}$, and $\delta^{TLS,2}_0=6.8 \times10^{-6}$. The other fitting parameters are volume of the TLS-inhabiting dielectrics $2 \mu m^3$, the asymmetry ratio $\eta=18$, and the exponent of the power dependent density of states of TLS $\mu=0.2$. We should note that this fit might be improved by including more species of TLS. However, this multiple-TLS model lacks a physical motivation and is not considered in this work.
\subsection{Power law fit}\label{alt_pwr_law}
Another well-known phenomenological fitting method introduces a free power law dependence to the saturation power term $1+\kappa$ \cite{Macha10}. The model when combined with the discrete formalism becomes 
\begin{eqnarray}
    \delta_\text{TLS} = \delta_0 \sum_n \frac{m_n\Gamma_2}{\Gamma_2^2(1+\kappa)^\beta+(\varepsilon_n/\hbar-\omega)^2} 
\end{eqnarray}
where $\beta$ is the exponent for the free power law dependence, and $\beta=1$ corresponds to the STM prediction. 
\begin{figure}
    \centering
    \includegraphics[width=0.9\textwidth]{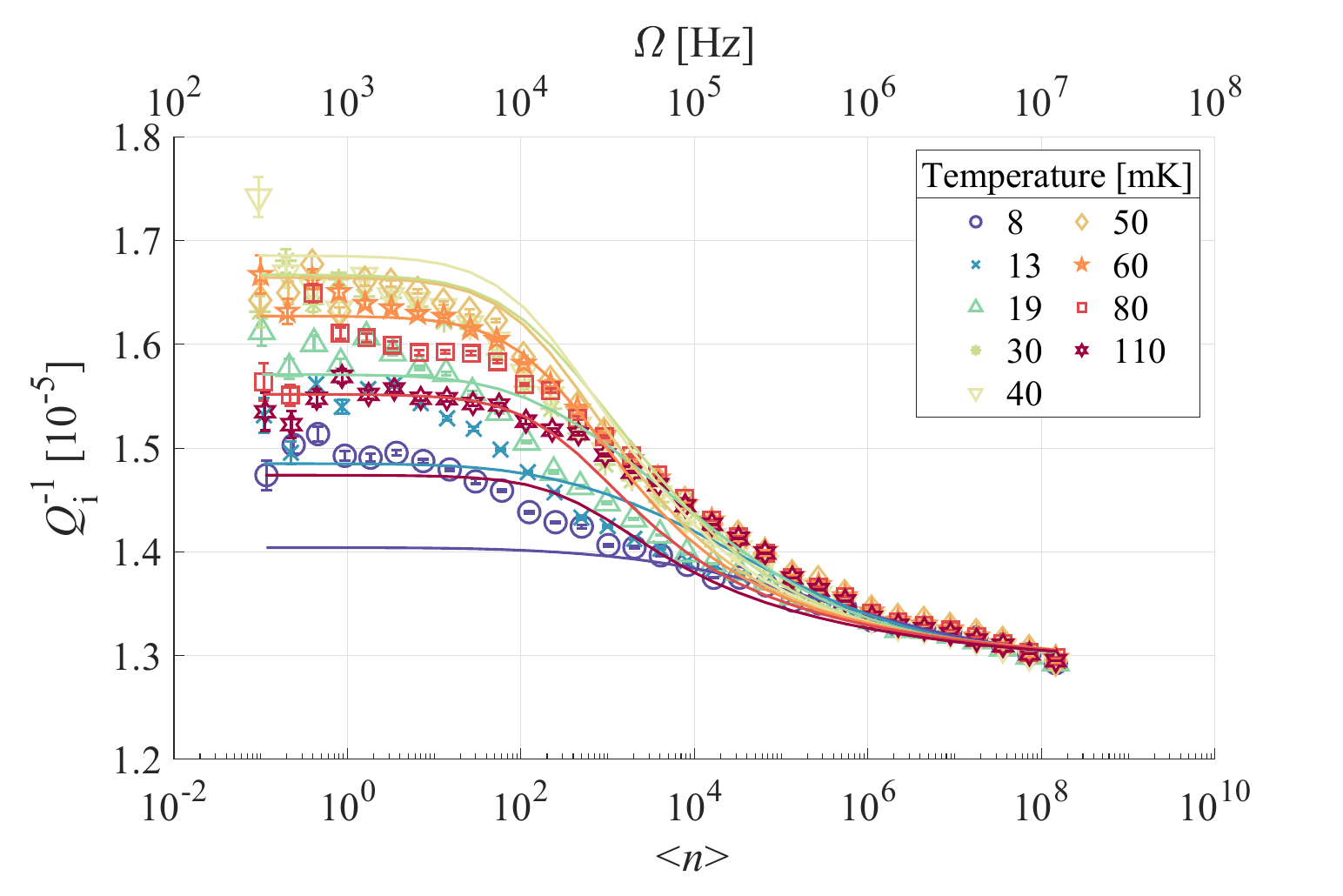}
    \caption{The discrete power law fit to the loss data  }
    \label{fig:alt_dispowl}
\end{figure}
The discrete free power law fit has a large deviation from the data as shown in Fig. \ref{fig:alt_dispowl}, with $\text{RMSE}=0.036$. The 8 fitting parameters are the power law exponent $\beta =0.38$, the exponent in the energy dependent density of states of TLS $\mu=0.2$, the volume of TLS-inhabiting dielectrics $2.6 \mu m ^3$, the  loss from other mechanisms $\delta_\text{other}=1.29\times10^{-5}$, the intrinsic TLS loss $\delta_0=8\times10^{-6}$, and the two relaxation rates are $\Gamma_1 = 700 \text{ to } 10^3 Hz$ and $\Gamma_2 = 3.3\times10^4 \text{ to } 1.4\times10^6 Hz$. 

Moreover, neither of the above alternative models provide a clear  interpretation of the physics in the TLS loss, since the phenomenological fitting parameters could have different microscopic origins, i.e. nonuniform field distribution, or different dipole moments or relaxation rates among several distinct ensembles of TLS. Not only does the discrete form of GTM in this work generate better fits to the measured loss, but it also provides a clear physical explanation for the observed loss: the low spectral density of coherent TLS and their interactions with the fluctuators are responsible for the observed temperature and power dependence.

\textcolor{black}{
\subsection{Fit with energy dependent density of states}\label{alt_nonzero_mu}
In Faoro and Ioffe's original work on GTM, an energy dependent density of states (DOS) for the TLS is assumed \cite{faoro_interacting_2015}, 
\begin{eqnarray}
    P(\Delta,\Delta_0)=(1+\mu)\left(\frac{\Delta}{\Delta_\text{max}}\right)^\mu\frac{P_0}{\Delta_0}
\end{eqnarray}
where $\mu\sim 0.3$ and $\Delta_\text{max}\sim \varepsilon_\text{max}\sim k_\text{B}(100K)$. The small exponent $\mu$ approximately characterizes the logarithmic reduction of the density of states of TLS at low energy in a system of interacting dipoles  \cite{cuevas_density_1989}. The energy dependent DOS will lead to a small correction in the linear temperature dependence of the dephasing rate, $\Gamma_\text{ds}\sim 10^{-3}(k_\text{B} T/\varepsilon_\text{max})^\mu k_\text{B} T/\hbar$. The model with an energy dependent density of states where $\mu$ is a free fitting parameter is used to fit to our data and is shown in Fig. \ref{fig:alt_mu}.\\
\begin{figure}
    \centering
    \includegraphics[scale=0.4]{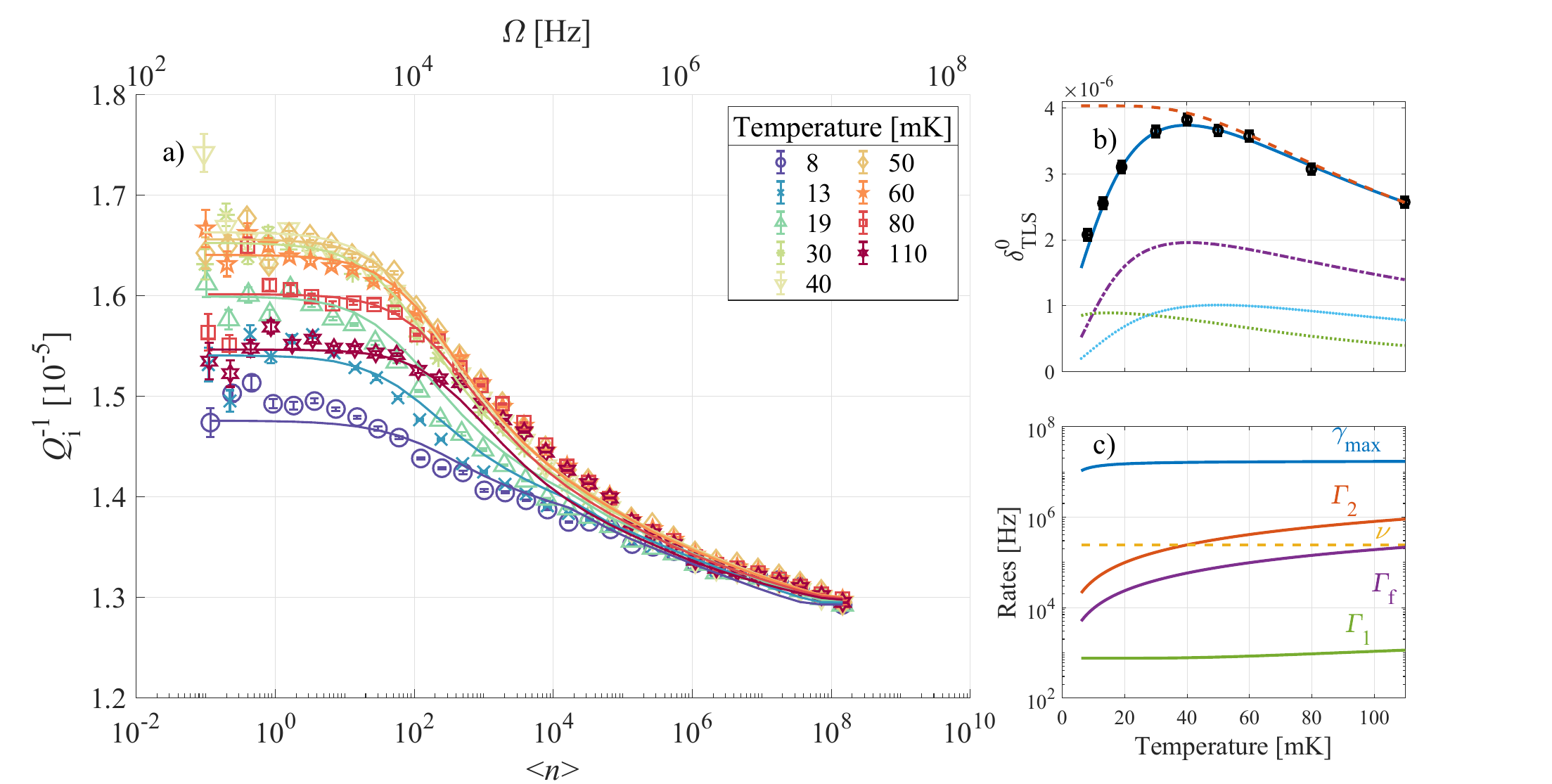}
    \caption{a) The least squares fit of the discrete GTM, \textcolor{black}{together with a constant background loss,} to the full power and temperature dependence of the measured internal loss below 150 mK. b) Plot of $\delta_\text{TLS}^0(T)$ extracted from the average of the low power loss below TLS saturation in Fig. \ref{Qi_T}.  The orange dashed curve is the temperature dependence of STM loss below saturation power $\propto\tanh{(\varepsilon/(2k_\text{B} T))}$. The purple dash-dotted (light blue densely dotted) curve is from the discrete summation of individual TLS contributions for low (intermediate)-$\gamma$ fluctuators at zero applied power. The green dashed curve is the temperature dependent low power limit of the TLS loss induced by high $\gamma$ fluctuators. The blue solid curve is the sum of contributions from the low, intermediate, and high $\gamma$ fluctuators. c) Comparison of the temperature dependent rates determined from the least squares fit. }
    \label{fig:alt_mu}
\end{figure}
The fit shows reasonable agreement with the data, with $\text{RMSE}=0.014$, slightly larger than that of the $\mu=0$ fit in the main text. There are in total 11 fitting parameters. The different contributions to the loss below TLS saturation power are plotted in Fig. \ref{fig:alt_mu} (b). Similar to the fit in the main text, the discrete TLS coupled to low and intermediate $\gamma$ fluctuators are responsible for the loss reduction. Different rates in the fit are summarized in Fig. \ref{fig:alt_mu} (c), with typical numerical values for TLS in amorphous materials \cite{faoro_interacting_2015}. The rates also satisfy $\Gamma_2 \gtrsim \Gamma_\text{f}$,  $\gamma_\text{max} \gg \Gamma_2$. In addition, the low temperature loss reduction occurs around $40$ mK as expected, when $\Gamma_2+\Gamma_\text{f} < \nu$, the width of the response is smaller than the detuning between TLS and the resonance.  The other quantities extracted from the fit are listed below: the volume of TLS-inhabiting dielectrics, $18.5\ \mu  m^3$, the intrinsic TLS loss, $\delta^{TLS}_0= 5.5\times10^{-6}$, the other loss, $\delta_\text{other}=1.29\times10^{-5}$, and the minimum fluctuator rate $\gamma_\text{min}=2.9\times10^{-4} \text{ Hz}$, and the exponent for the energy dependent density of states $\mu=0.2$.
}

\section{Non-Equilibrium Quasiparticle Treatment and \texorpdfstring{$n_\text{qp}$}{nqp} and \texorpdfstring{$Q_\text{qp}$}{Qqp} Estimates}\label{nonep_qp_sup}
The aim of this section is to examine the non-equilibrium quasiparticle density by simulating the physics using the non-equilibrium energy-dependent distribution of quasiparticles. This model (scattering model) was established by considering the coupled quasiparticle and phonon systems.\cite{ChangScal77} The details of this model are described in references \cite{Goldie} \cite{Visser2014} and \cite{Budoyo2015} and employ numerical methods to discretize the distribution of quasiparticles $f(E)$ at energy $E$ and the phonon distribution $n(\Omega)$ at energy $\Omega$ by solving the kinetic equations in steady state for a given flux of microwave photons and higher frequency radiation.

\begin{figure} [ht]
\centering
\includegraphics[width=0.48\textwidth]{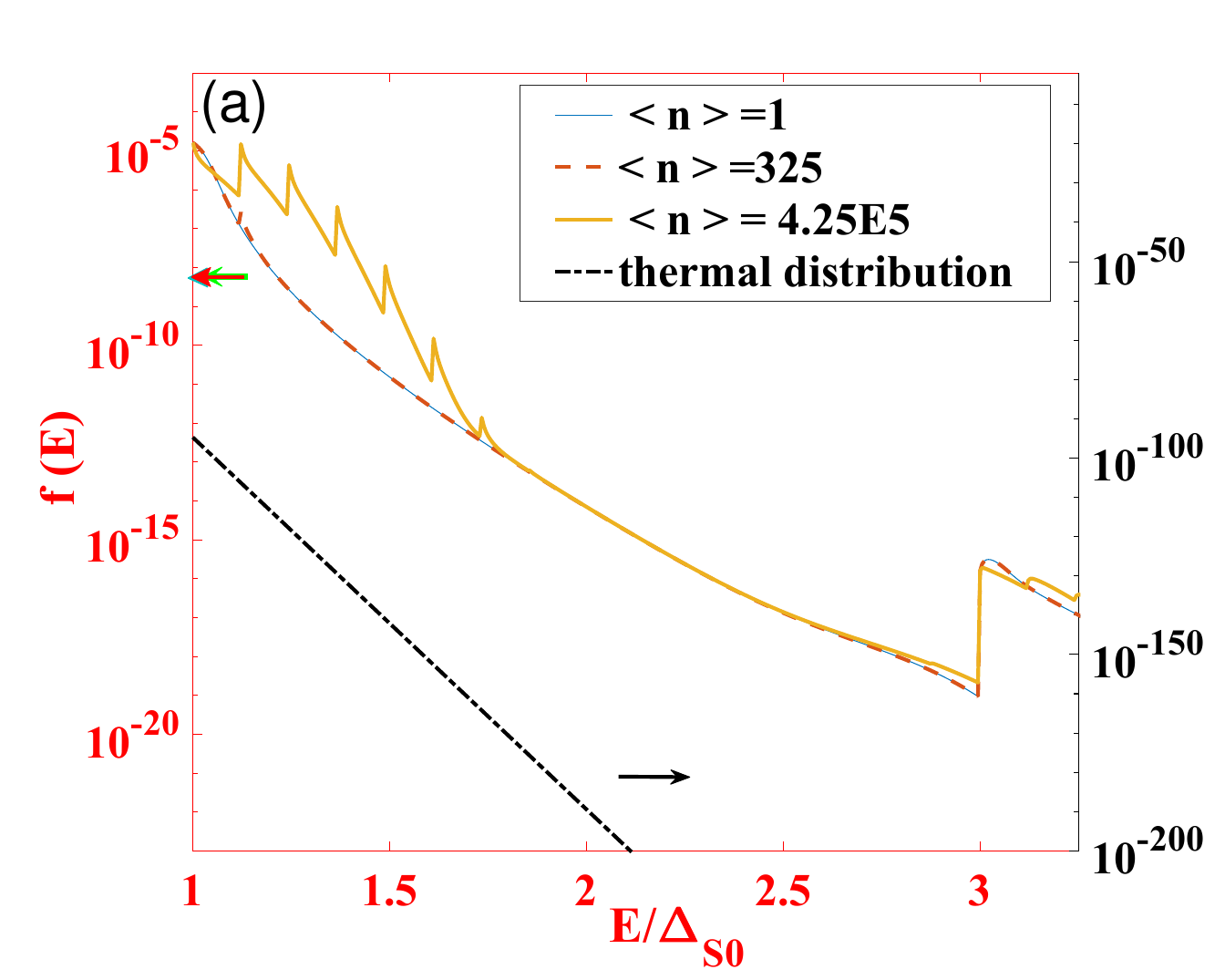}
\includegraphics[width=0.46\textwidth]{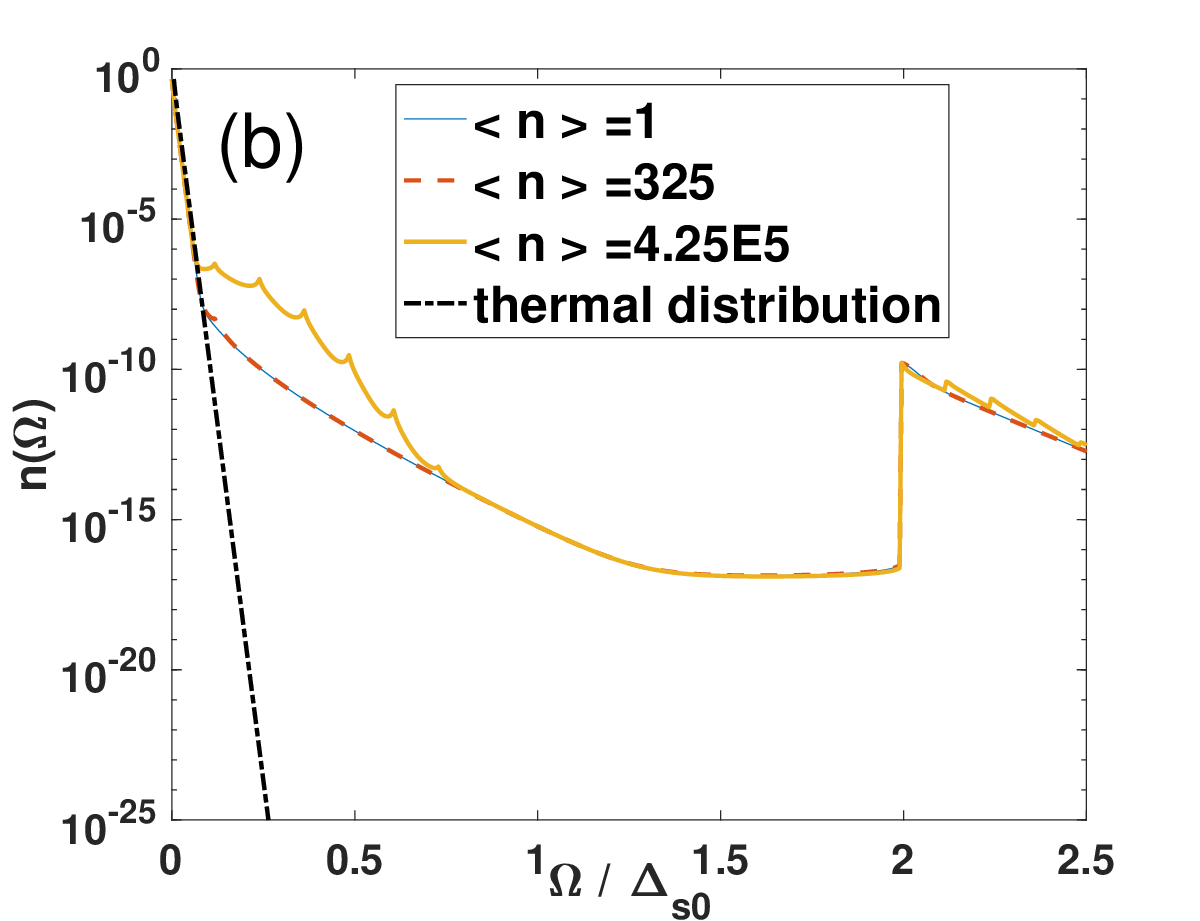}
\caption{\label{f_n_E} The calculated quasiparticle distribution $f(E)$ (a) and phonon distribution $n(\Omega)$ (b) as a function of normalized energy for different circulating numbers of photons in the half wavelength resonator. Note that in the plot of $f(E)$, a double y axis is used due to the different scales of thermal and non-equilibrium distributions.}
\end{figure}

Fig. \ref{f_n_E} (a) and Fig. \ref{f_n_E} (b) show the calculated quasiparticle distribution and phonon distribution, respectively, for three selected circulated numbers of photons in the half wavelength resonator. This calculation is performed by assuming the fridge base temperature $T_\text{b}= 10 \: mK$, the resonator drive frequency 3.6442 GHz ($\hbar \omega = 23 \: \mu eV$), superconducting energy gap $\Delta_{\text{S}0}=188$ $\mu eV$, $Q_\text{i}=10^5$, $Q_\text{c}=1.5\times 10^6$ which we obtain from fitting the resonance, resonator center conductor volume $8.6\times 10^{-14}$ $m^{3}$ \ and effective temperature $T_\text{eff}=189 \: mK$ due to stray light illumination and radiation which creates enhanced phonon generation, as described by the Parker model \cite{Parker_supporting}. Table \ref{Tab:scattering} lists all of the parameter values used in this model. The black dashed line in Fig. \ref{f_n_E}(a) indicates the thermal distribution of quasiparticles without any microwave excitation at $T_\text{b}= 10$ $mK$.  At low circulating photon numbers ($\langle n \rangle=1$ or 325 ), the quasiparticle distribution is enhanced significantly above the $T_\text{b}$ thermal distribution, and the phonon distribution is also enhanced. At high circulating photon numbers, jumps appear in the electron and phonon distributions every $\hbar \omega$ because microwave drive at high power significantly affects the distributions. Big step jumps at $E=3 \Delta_{\text{S}0}$ and $\Omega=2 \Delta_{\text{S}0}$ are due to pair breaking and recombination processes.

From the quasiparticle distribution $f(E)$, one can calculate the quasiparticle density, $n_\text{qp}$, by numerical integration over all energy, with its density of states $\rho(E)$
\begin{equation}
n_\text{qp}=4N_0 \int_{\Delta_{\text{S}0}}^{\infty} f(E) \rho(E) dE \quad \text{with} \quad \rho(E)=\frac{E}{\sqrt{E^2-\Delta_{\text{S}0}^2}} 
\label{nqp_eq}
\end{equation}

\begin{figure} [ht]
\includegraphics[width=0.47\textwidth]{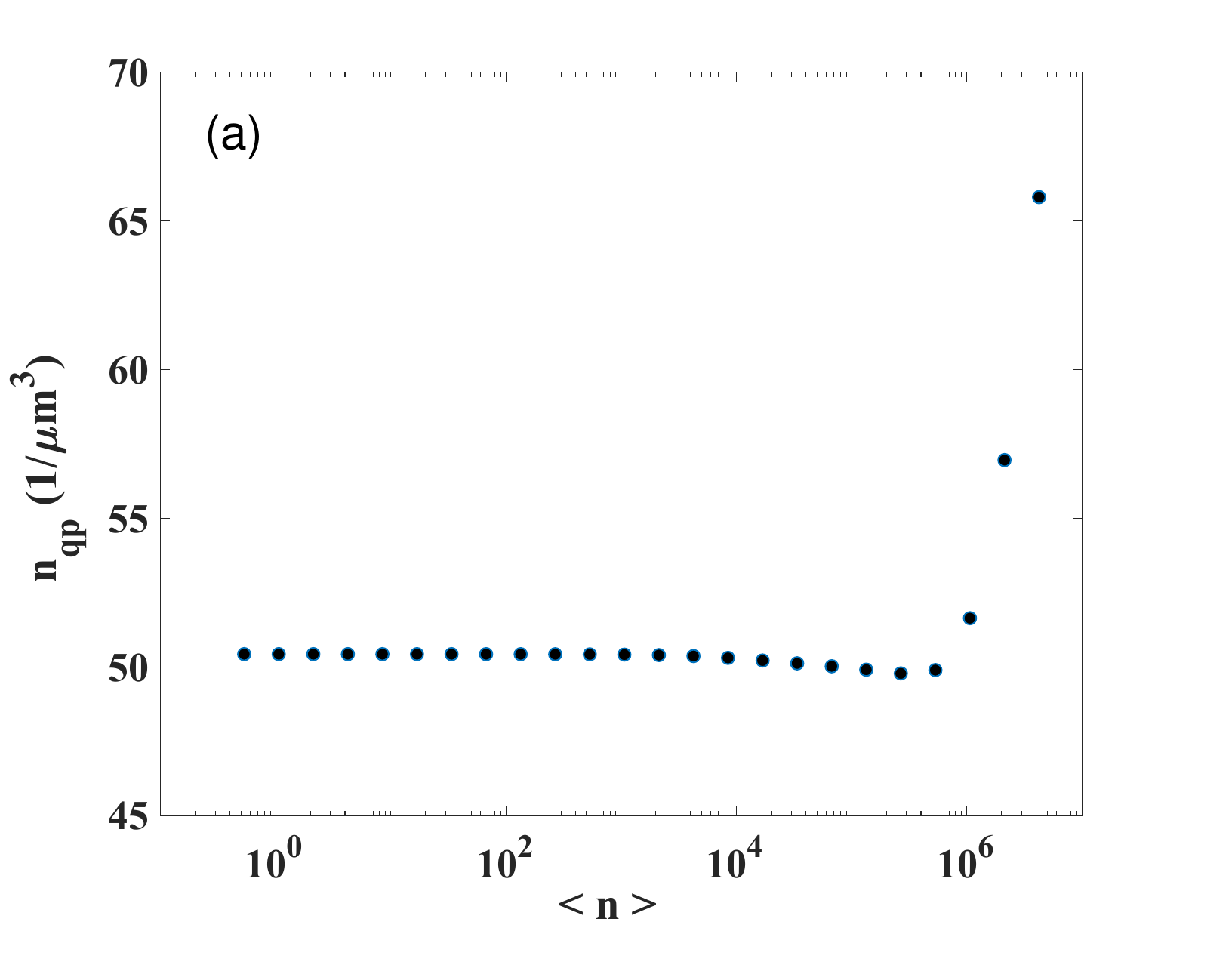}
\includegraphics[width=0.45\textwidth]{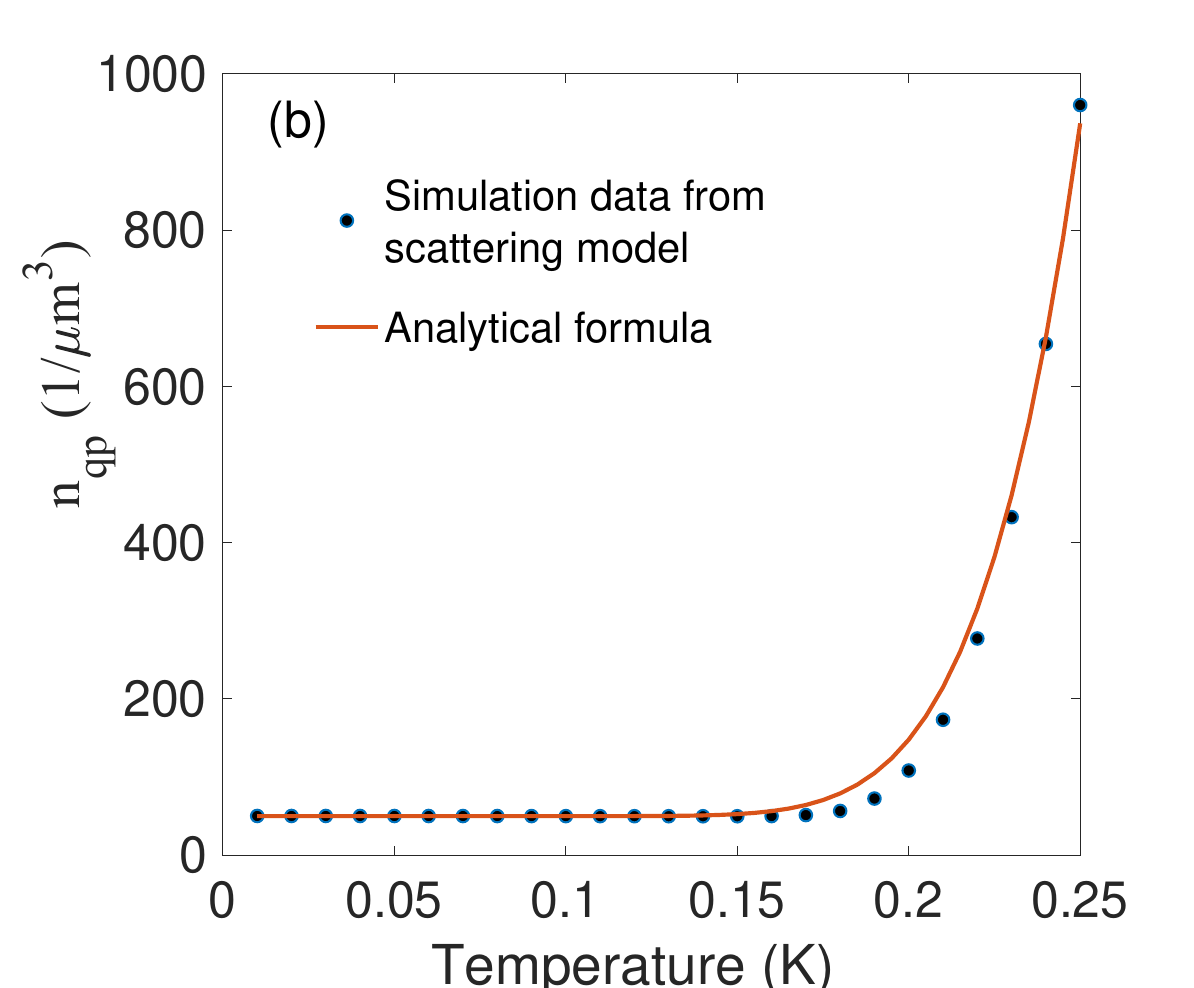}
\caption{\label{Nqp} (a) The calculated quasiparticle density as a function of circulating photon numbers in the half wavelength resonator. Here $T_\text{b}=10 \: mK$ and $T_\text{eff}=189 \: mK$. (b) The calculated quasiparticle density as a function of fridge temperature at $\langle n \rangle=1$. Note the dots are calculated from the non-equilibrium model, and the solid line is from Eq. (\ref{n_qp_analyticEq}) of the main text. }
\end{figure}

The calculated $n_\text{qp}$ at different circulated photon numbers $\langle n \rangle$ are shown in Fig. \ref{Nqp}(a). Because this calculation is performed at the 10 mK fridge base temperature, the entire term $n_\text{qp}$ (Eq. (\ref{n_qp_analyticEq}) of the main text) can be regarded as the contribution of non-equilibrium quasi-particle density since the number of thermal quasiparticles is extremely small.  The calculated $n_\text{qp}$ is around  $50 \; \mu m^{-3}$ when $\langle n \rangle \; < 10 ^6$ .  In addition, the calculated result of $n_\text{qp}$ as a function of fridge/bath temperature ($T_\text{b}$) at $\langle n \rangle=1$ is shown in Fig. \ref{Nqp}(b) as circle dots. When the bath temperature is below 150 $mK$, $n_\text{qp}$ remains constant. This is consistent to the assumptions made in the frequency shift fit at low temperature where quasiparticle density is a constant. The continuous line of Fig. \ref{Nqp}(b) is from Eq. (\ref{n_qp_analyticEq}) of the main text.

For the quality factor due to the quasiparticles, $Q_\text{qp}$ is defined as 
\begin{equation}
 \frac{1}{Q_\text{qp}}=\alpha \frac{\sigma_1}{\sigma_2}
\label{Qqp_eq}
\end{equation}
where the kinetic inductance ratio $\alpha=0.0185$ was obtained from the frequency shift fit in the main text. The real and imaginary parts of the complex conductivity $\sigma =\sigma_1-j\sigma_2$ can be expressed respectively by the Mattis-Bardeen formula \cite{Mattis-Bardeen_supporting} given by

\begin{eqnarray}
\frac{\sigma_1}{\sigma_\text{N}}(\omega)=\frac{2}{\hbar \omega} \int_{\Delta_{\text{S}0}}^{\infty} \left[ f(E)-f(E+\hbar \omega) \right] g_1(E) dE 
\\
\frac{\sigma_2}{\sigma_\text{N}}(\omega)=\frac{2}{\hbar \omega} \int_{\Delta_{\text{S}0}-\hbar \omega}^{\infty} \left[ f(E)-2f(E+\hbar \omega) \right] g_2(E) dE 
\label{conductivity_eq}
\end{eqnarray}

where $\sigma_N$ is the normal-state conductivity, $g_1(E)=h_1(E,E+\hbar \omega) \rho (E)$ and $g_2(E)=h_1(E,E+\hbar \omega) \frac{E}{\sqrt{\Delta_{\text{S}0}^2-E^2}}$ with $h_1(E,E')=(1+\frac{\Delta_{\text{S}0}^2}{E*E'})\rho(E'). $  Here we use the non-equilibrium distribution function $f(E)$ discussed above.
\\

\begin{figure} [ht]
\centering
\includegraphics[width=0.5\textwidth]{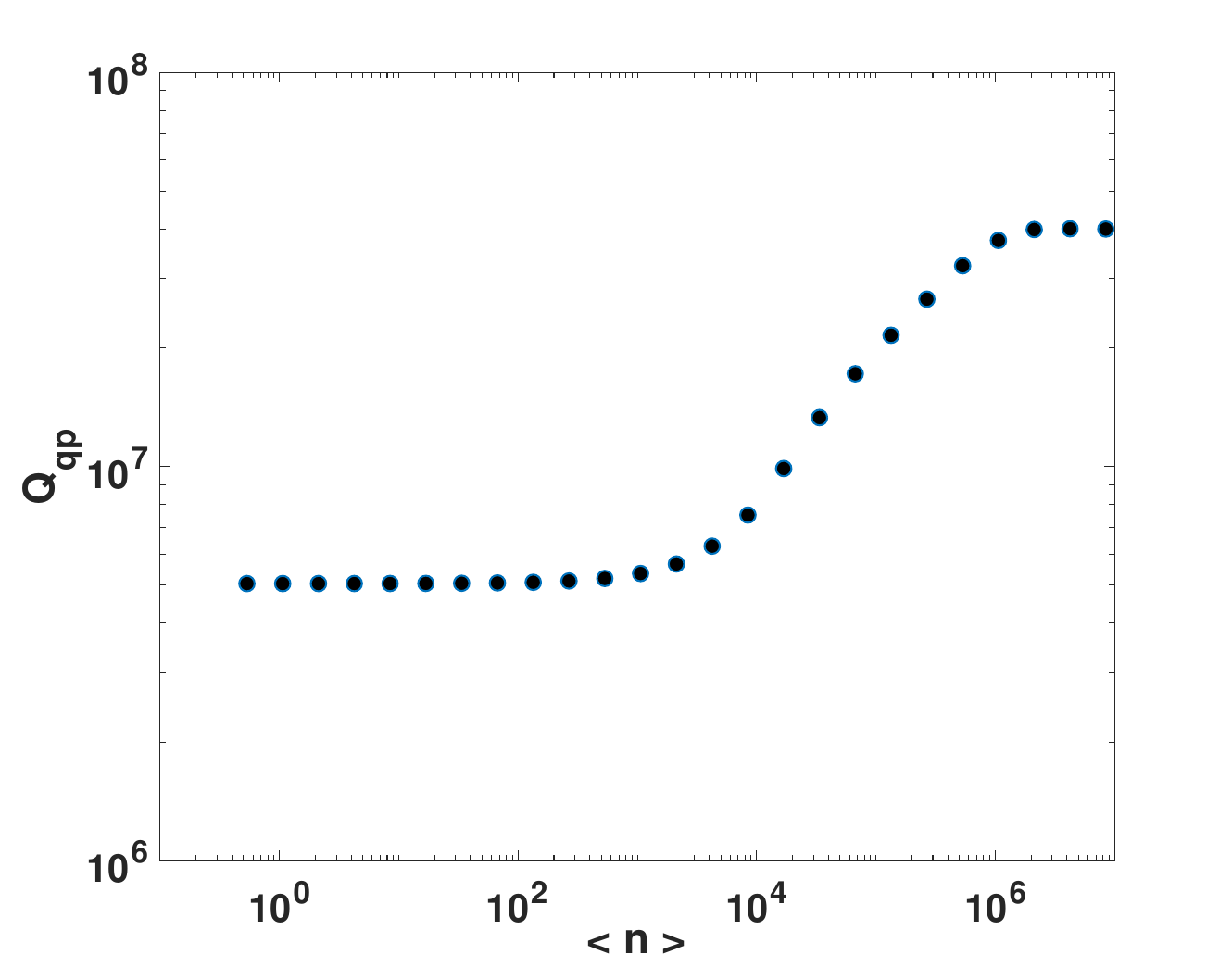}
\caption{\label{Q_qp} The calculated quality factor of quasi-particle  as a function of circulating photon numbers $\langle n \rangle$ in the half wavelength resonator. Here $T_\text{b}=10 \: mK$ and $T_\text{eff}=189 \: mK$, and other parameters are listed in Table \ref{Tab:scattering}.}
\end{figure}

The calculated result of $Q_\text{qp}$ as a function of average circulated photon numbers is shown in Fig. \ref{Q_qp}. In the regime of low circulating photon numbers, $Q_\text{qp}$ remains a constant and then gradually increases as the circulating photon numbers in the half wavelength resonator increase.  Overall, the $Q_\text{qp}$ is on the order of $10^7$, and this result verifies our assumption that $Q_\text{qp} > Q_\text{TLS0}$ in the $Q_\text{i}(T)$ fitting. In other words, the loss in the half wave length resonator at low temperatures and low circulating photon numbers is dominated by the TLS losses.  

\begin{table}
\centering
\begin{tabular}
{| >{\raggedright\arraybackslash}p{0.4\linewidth} | >{\raggedright\arraybackslash}p{0.4\linewidth} |}
\hline
Superconducting gap ($\Delta_{\text{S}0}$) = $188 \: \mu eV $  &  RF photon energy ($\hbar \omega $) = $23 \:\mu eV $  \\
    & \\ 
$Q_\text{i}=1.0 \times10^5$  & $Q_\text{c}=1.5\times10^6 $ \\
    & \\
Quasiparticle-phonon time =438 ns \cite{Visser2014} &  Characteristic phonon time = 0.26 ns \cite{Visser2014}\\
    & \\
Phonon escape time =0.17 ns \cite{Visser2014}  & Resonator volume= $8.6 \times 10^{-14} \; m^3$ \\
    & \\
Base Temperature $T_\text{b}$=10 mK     & Phonon effective temperature ($T_\text{eff}$)=189 mK \\
\hline
\end{tabular}
\caption{\label{Tab:scattering} Parameters used for our non equilibrium quasi-particle calculation.}
\end{table}

\section{CST Microwave Simulation of CPW Electric Fields}\label{CST_section}
\begin{figure}[ht]
    \centering
    \includegraphics[width=170 mm]{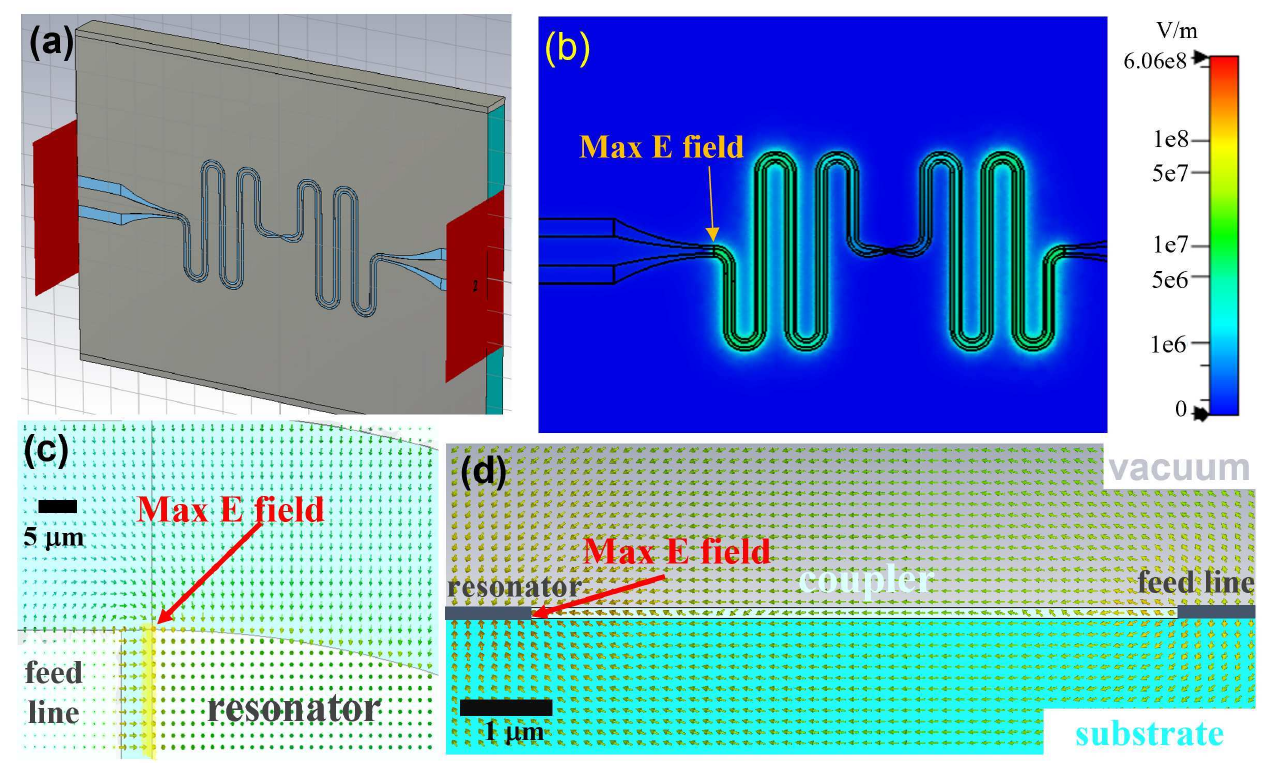}
    \caption{CPW resonator model in CST microwave studio. (a) The geometry of the CPW resonator model in CST. The coupling capacitors and the entire resonator structure reproduce those  in the experiment. (b) The top view of the E field strength on the substrate-vacuum interface at the fundamental resonance of 3.647 GHz, which has a node in the center of the resonator, a typical standing wave pattern in a half wavelength resonator. (c) The close-up top view of the E field vector plot on the substrate-vacuum interface near the coupling capacitor. The maximum E field is found around the corner of the center strip that is part of the coupling capacitor.The E field that contributes to the TLS model is estimated from the average of the field along the resonator side of the coupler about $1.16\times10^8 Vm^{-1}$, highlighted in the yellow-orange line.  (d) The side view of the E field vector plot on the cross section through the coupler at the center strip corner. The maximum is at the corner of the center strip on the substrate-vacuum interface.}
    \label{fig:cst}
\end{figure}

A model of the CPW microwave resonator was constructed in CST Microwave Studio. The model structure (Fig. \ref{fig:cst}) represents the entire CPW resonator and coupling capacitors, both of which reproduce the geometrical structure in our superconducting chip. The superconductor is modeled as a perfect electric conductor for the purpose of E field calculation. The results below are obtained from the finite element frequency domain solver in CST. 

The CST simulation shows that on resonance at 3.647 GHz, the electric field can attain a maximum of $6\times 10^8$ V/m on the substrate-vacuum interface at the corner of the center strip that is part of the coupling capacitor on the input side. This calculation was done under a 0.5 watt excitation level. An average electric field of the adjacent area is estimated to be $1.16\times10^8$ $V/m$. By scaling this power down to that required to achieve one circulating photon in the resonator, we estimate the average electric field of the region nearby the coupling capacitor of the resonator to be $0.2$ V/m.  In Fig. \ref{Qi_T} and \ref{fig:delta0_pc}, the Rabi frequency $\Omega$ on the x-axis is estimated from this electric field and a dipole moment estimated from \cite{Sarabi16_supporting,Chih22}.  

\end{document}